\documentclass{aa}
\usepackage{graphicx}
\usepackage{txfonts}
\usepackage{natbib}

\begin{document}

   \title{FIRST-based survey of Compact Steep Spectrum sources}

   \subtitle{IV. Multifrequency VLBA observations of very compact objects}

   \author{M. Kunert-Bajraszewska\inst{1}
          \and A. Marecki\inst{1}
          \and P. Thomasson\inst{2}
          }

   \authorrunning{Kunert-Bajraszewska et al.}

   \offprints{Andrzej Marecki\\
              \email{amr@astro.uni.torun.pl}}

   \institute{Toru\'n Centre for Astronomy, N. Copernicus University,
              87-100 Toru\'n, Poland
         \and
              Jodrell Bank Observatory, University of Manchester,
              Macclesfield, Cheshire, SK11 9DL, UK
                }

   \date{Received 27 October 2005/ Accepted 7 January 2006}

\abstract
{Evidence has been mounting recently that activity in some radio-loud
AGNs (RLAGNs) can cease shortly after ignition and that perhaps even a majority
of very compact sources may be short-lived phenomena because of a lack of stable
fuelling from the black hole. Thus, they can fade out before having evolved to
large, extended objects. Re-ignition of the activity in such objects is not
ruled out.} 
{With the aim of finding more examples of these objects and to investigate
if they could be RLAGNs switched off at very early
stages of their evolution, multifrequency VLBA observations of six sources with
angular sizes significantly less than an arcsecond, yet having steep
spectra, have been made.}
{Observations were initially made at 1.65\,GHz using the VLBA with the
inclusion of Effelsberg telescope. The sources were then re-observed with the
VLBA at 5, 8.4 and 15.4\,GHz. All the observations were carried out in a
snapshot mode with phase referencing.}
{One of the sources studied, 0809+404, is dominated by a compact component
but also has diffuse, arcsecond-scale emission visible in VLA images.
The VLBI observations of the ``core'' structure have revealed that this
is also diffuse and fading away at higher frequencies.
Thus, the inner component of 0809+404 could be a compact fading object.
The remaining five sources presented here show either core-jet or
edge-brightened double-lobed structures indicating that they are in an
active phase.}
{The above result is an indication that the activity of the host galaxy of
0809+404 may be intermittent. Previous observations obtained from the literature
and those presented here indicate that activity had ceased once in the past,
then restarted, and has recently switched off again.}


\maketitle


\section{Introduction}

The activity period for radio-loud AGNs (RLAGN) can last up to $\sim$$10^{8}$
years \citep{al87,liu92} and, as their lobes are huge reservoirs
of energy, even if the energy supply from the central engine to the
hotspots and the lobes eventually cuts off, the radio sources are still
observable for a substantial period of time. This so-called ``coasting
phase'' of the lobes of a RLAGN can last up to $10^{8}$\,yr
\citep{kom94,sl01} and preserves information of past nuclear activity.
As the source gradually fades out, its spectrum becomes steeper and
steeper because of radiation and expansion losses. Objects possessing
these features are sometimes termed ``relics'' or ``faders''.

The structure and other properties of the double radio source B2\,0924+30,
identified with an E/S0 galaxy IC\,2476, shows it to be a good example of a
fader \citep{cor87}. The projected linear size of the whole system is
270\,$h^{-1}$\,kpc\footnote{For consistency with earlier papers in this field,
the following cosmological parameters have been adopted throughout this paper:
$H_0$=100${\rm\,km\,s^{-1}\,Mpc^{-1}}$ and $q_0$=0.5. Wherever in the text the
linear sizes are referred to, $h^{-1}$ is introduced.} which indicates that it
is a Large Symmetric Object (LSO). \citet{jam04} have confirmed that
B2\,0924+30 is indeed a relic radio structure and that it switched off its
activity $\sim$$5\times10^7$ years ago and, as such, can be labelled a ``dead''
radio galaxy. A large number of examples of fossil radio galaxies or cluster
relic systems have recently been found by \citet{coh04} as a result of their
VLA observations at 74\,MHz.

There are no obvious reasons for the
existence of a lower limit to the length of the activity period in a RLAGN;
it could be shorter than has already been seen for LSOs. If this is
the case it could be that the growth of the radio source has been impeded, even
at an early stage of its evolution. As a result, small-scale faders might
exist. An attempt to test observationally if there are young faders among
Medium-sized Symmetric Objects (MSOs), i.e. the objects that have linear
sizes in the range 1--20 $h^{-1}$\,kpc, has been made by \citet{kmts05} --
hereafter Paper~II. This revealed one strong candidate --~1542+323. 

\citet{gir05} have described a class of low power compact (LPC) radio
sources, their small sizes and moderate luminosities (comparable to those of
low power giant FR\,I radio galaxies) being ascribed to a number of different
physical reasons: youth, low kinetic power of the jets or frustration and also
the premature end of nuclear activity. 1855+37, one of the sources they
investigated, could be a very good example of a compact fader.

There is no reason why one should not search for faders among the
most compact of radio sources, particularly those belonging to the class of
Compact Symmetric Objects (CSOs), i.e. sub-kiloparsec-scale extragalactic radio
sources with symmetric radio structures. Most of these sources are triples with
the central component being a radio core, or doubles with only two detectable
radio lobes. Some CSOs have radio spectra that peak at a few
gigahertz and have been classified as Gigahertz-Peaked Spectrum (GPS)
sources. GPS radio galaxies seem to be CSOs with a simple structure
\citep{odea98}. However, most GPS quasars with a core-jet or complex
milliarcsecond-scale morphology do not seem to have symmetric structures and
they are thought to be a different class of object \citep{stan03}.

The physical origin of CSOs has been explained in two ways. According to the
frustration scenario \citep{bmh84} the radio source is permanently confined to
a region within the host galaxy by a dense environment from which it is
impossible for it to evolve into a large source. \citet{alex00} also claims
that the observational density of sources in the power--linear size plane can
be reproduced only if it is assumed that there exists a class of young
frustrated objects. They would remain very weak and have diffuse hotspots
and a brighter jet. Those objects, which manage to escape the high-density
regions, would continue their evolution and reach the classical FR\,II stage.

Alternatively, \citet{pm82} and \citet{c85} suggested that CSOs -- they were
labelled compact-doubles at that time; the term ``CSO'' was introduced later by
\citet{wil94} and \citet{r94} -- could be young radio sources
that would evolve into large radio objects during their lifetimes. Based upon
this scenario, \citet{r96} proposed an evolutionary scheme unifying three
classes of radio sources: CSOs, MSOs -- a subset of the Compact Steep Spectrum
sources (CSS) class -- and LSOs.

At the present time this is the youth scenario that is generally accepted;
see the review by \citet{fanti00}. An argument in favour of this has been
found mainly in age measurements of individual classes of radio sources: CSOs
are younger than $\sim$$10^4$~years \citep{ocp98,gir03}, MSOs are typically
$\sim$$10^5$~years old \citep{mur99} and LSOs can manifest their activity for
up to $\sim$$10^8$~years \citep{al87,liu92}.
\citet{sn99,sn00} added an important
ingredient to this scheme, namely that the radio luminosities of CSOs increase
as they evolve, reach a maximum in the MSO phase and then gradually decrease as
these objects increase in size to become LSOs.

\citet{mar46,mks06} claim that this evolutionary track is not the only one
possible. In fact, several, if not a ``continuum'' of such tracks might exist
and the one shown by \citet{sn99,sn00} just {\em appears} as the only one
simply because of selection effects. If the energy supply cuts off early, the
object leaves the ``main sequence'' proposed by \citet{sn99,sn00} and will
never reach the LSO stage, at least in a given phase of activity. Thus, there
should exist a class of (very) small-scale objects that resemble large-scale
faders. This conforms to early predictions that many CSOs could be young
objects that switch off after a short period of time \citep{r94}. Strong
support for such an idea also comes from \citet{rb97} who proposed a model in
which extragalactic radio sources are intermittent on timescales of
$\sim$$10^4$--$10^5$~years. The above considerations have recently been
supported by observational results obtained by \citet{gug05}, suggesting that
many CSOs die young or are episodic in nature, and so it is likely that only
a minority of them ``survive'' and further evolve.

In this paper -- the fourth of the series -- VLBA observations of 6~compact
sources which are candidates for prematurely dying CSOs are presented and
discussed.

\begin{table*}[t]
\caption[]{Optical magnitudes and radio flux densities of target sources at
two frequencies.}
\begin{center}
\begin{tabular}{@{}c c c c l l c l c l r l@{}}
\hline
\hline
       &    &     &   &     &   &
\multicolumn{1}{c}{Total} &
&
\multicolumn{1}{c}{Total} &     &  &  \\
~~~Source & RA & Dec & ID&
\multicolumn{1}{c}{$m_{R}$}&
\multicolumn{1}{c}{\it z}&
\multicolumn{1}{c}{flux at}&
\multicolumn{1}{c}{log$P_{1.4\mathrm{GHz}}$}&
\multicolumn{1}{c}{flux at}&
\multicolumn{1}{c}{$\alpha_{1.4\mathrm{GHz}}^{4.85\mathrm{GHz}}$}&
\multicolumn{1}{c}{LAS}&
\multicolumn{1}{c}{LLS}\\
~~~Name   & h~m~s & $\degr$~$\arcmin$~$\arcsec$ &  & & &1.4\,GHz&
&4.85\,GHz&&&\\
       &    &     &   &     &   &
\multicolumn{1}{c}{[mJy]}&
\multicolumn{1}{c}{[W~Hz$^{-1}$]} &
\multicolumn{1}{c}{[mJy]}& &
\multicolumn{1}{c}{[mas]}&
[$h^{-1}~{\rm pc}]~~~$ \\
~~~(1)& (2)& (3) &(4)&
\multicolumn{1}{c}{(5)}&
\multicolumn{1}{c}{(6)}&
\multicolumn{1}{c}{(7)}&
\multicolumn{1}{c}{(8)}&
\multicolumn{1}{c}{(9)}&
\multicolumn{1}{c}{(10)}&
\multicolumn{1}{c}{(11)}&
\multicolumn{1}{c}{(12)}\\
\hline
~~~0809+404 &08 12 53.131 &40 19 00.09&Q&19.50&~0.551&1068&~~~26.63& 392&
$-$0.81&14.87&~~~~54.9\\
~~~0949+287 &09 52 06.090 &28 28 32.35&G&20.15&~~~~-- &1364&~~~27.38$\ast$& 529&
$-$0.76&308.79&1330.0$\ast$\\
~~~1159+395 &12 01 49.982 &39 19 11.26&G&23.32&~2.370&~~599&~~~27.78& 249&
$-$0.71&41.08&~~161.3\\
~~~1315+396 &13 17 18.653 &39 25 28.02&Q&18.20&~1.560&~~615&~~~27.39& 227&
$-$0.80&33.77&~~143.8\\
~~~1502+291 &15 04 26.715 &28 54 30.55&q&18.58&~0.056?&~~567&~~~24.29?& 261&
$-$0.63&41.60&~~~~30.8?\\
~~~1616+366 &16 18 23.546 &36 32 01.33&G&19.35&~0.734&~~536&~~~26.60&268&
$-$0.56&$\sim$~60.00&~~242.0\\
\hline
\end{tabular}
\end{center}

\vspace{0.5cm}
{\small
Description of the columns:
\begin{itemize}
\item[---]{Col.~~(1): Source name in the IAU format;}
\item[---]{Col.~~(2): Source right ascension (J2000) extracted from FIRST;}
\item[---]{Col.~~(3): Source declination (J2000) extracted from FIRST;}
\item[---]{Col.~~(4): Optical identification: G - galaxy, Q - quasar, q -
star-like object without known redshift;}
\item[---]{Col.~~(5): Red magnitude extracted from SDSS/DR4;}
\item[---]{Col.~~(6): redshift;}
\item[---]{Col.~~(7): Total flux density at 1.4\,GHz extracted from FIRST;}
\item[---]Col.~~(8): Log of radio luminosity at 1.4\,GHz in
W~Hz$^{-1}$;
\item[---]{Col.~~(9): Total flux density at 4.85\,GHz extracted from GB6;}
\item[---]{Col.~~(10): Spectral index between
1.4 and 4.85\,GHz calculated using flux densities in columns (7) and (9);}
\item[---]{Col.~(11): Largest Angular Size (LAS) in milliarcseconds measured
in the 1.65-GHz VLBA image --- in most cases, as a separation between the
outermost component peaks; in one case (denoted with``$\sim$'') measured in the
image contour plot;}
\item[---]{Col.~(12): Largest Linear Size (LLS) in $h^{-1}$~pc.}

Values denoted with ``$\ast$'' were computed assuming median redshift
$z=1.085$.\\
Redshift quoted for 1502+291 is the redshift of Abell\,2022 cluster.
\end{itemize}
}

\label{table1}
\end{table*}

\section{The observations and data reduction}

A sample of 60~candidates which could be weak Compact Steep Spectrum
sources was selected from the VLA FIRST catalogue
\citep{wbhg97}\footnote{Official website: http://sundog.stsci.edu}.
The selection criteria have been given by \citet{kun02} -- hereafter Paper~I.
All the sources were initially observed with MERLIN at
5\,GHz and the results of these observations led to the selection of several
groups of objects for further study with MERLIN and the VLA (Paper~II), as well
as the VLBA and the EVN \citep[][hereafter Paper~III]{mks06}. One of those
groups contained six sources that were barely resolved by MERLIN at 5\,GHz, but
still had steep ($\alpha \leq -0.5, S\propto\nu^{\alpha}$) spectra between
1.4 and 4.85\,GHz.
The basic properties of the six sources are given in Table~\ref{table1}.

Initial 1.65-GHz VLBA observations of the sources listed in Table~\ref{table1}
were carried out on 27 and 28 July 2002 in a snapshot mode with
phase referencing. The Effelsberg telescope was also included in order to
improve the resolution at that comparatively low frequency. Each target source
scan was interleaved with a scan on a phase reference source throughout an
$\sim$8-hour track. The total cycle time (target and phase reference) was
9~minutes including telescope drive times, with $\sim$7\,minutes actually on the
target source per cycle.

After a careful inspection of the 1.65-GHz images, it was decided that all
the sources except one should be observed at 5, 8.4 and 15.4\,GHz. The
exception was 1502+291, for which it was considered that only an observation
at 5\,GHz would be necessary to confirm its core-jet structure, 
indicating that it was not a candidate for a ``switched off'' object. These
follow-up VLBA-only observations were carried out on 13 and 14 October 2003.
(Only 9~telescopes were used for the observations due to the failure of 
VLBA-KP antenna.) Each target source scan was
interleaved with a scan on a phase reference source throughout an $\sim$12-hour
track. The total cycle time (target and phase reference) was 8~minutes
including telescope drive times, with $\sim$6\,minutes actually on the target
source per cycle at each frequency. Except for 1502+291, the cycles for a given
target-calibrator pair were grouped and rotated round the three frequencies.
Only two sources (1315+396, 1616+366) have been detected at 15.4\,GHz.

\begin{figure*}[t]
\centering
\includegraphics[width=7cm, height=7cm]{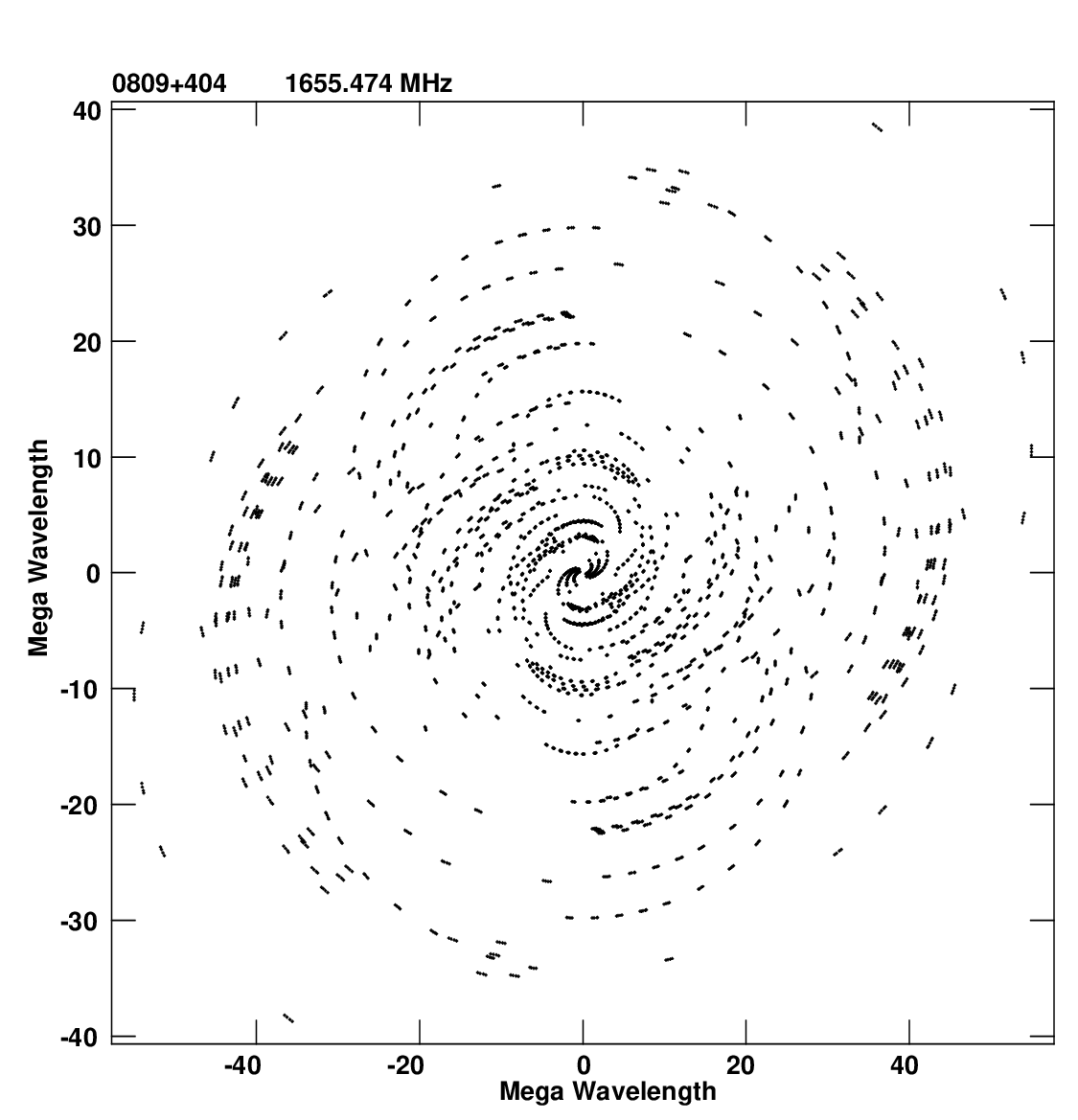}
\includegraphics[width=7cm, height=7cm]{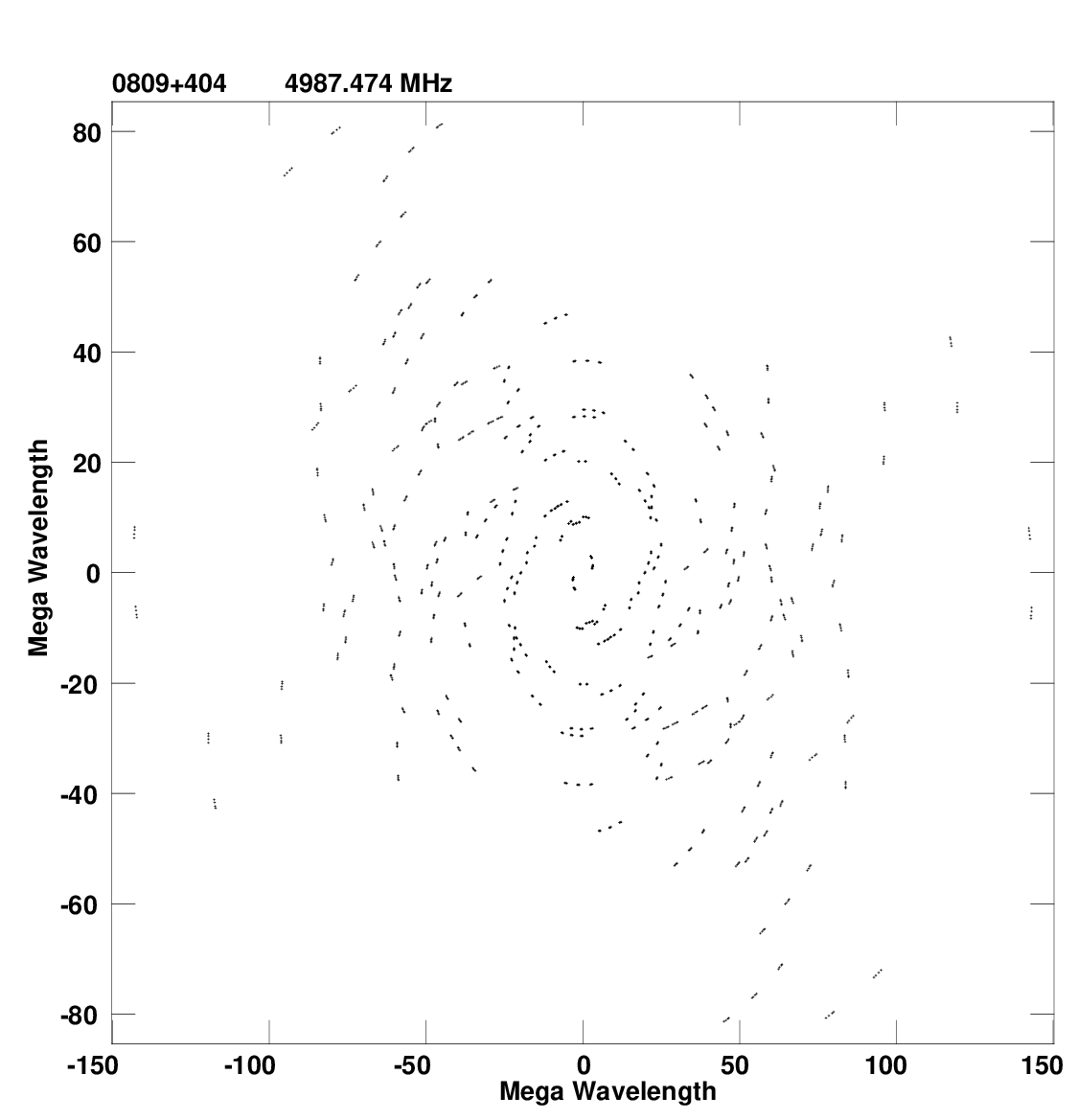}
\includegraphics[width=7cm, height=7cm]{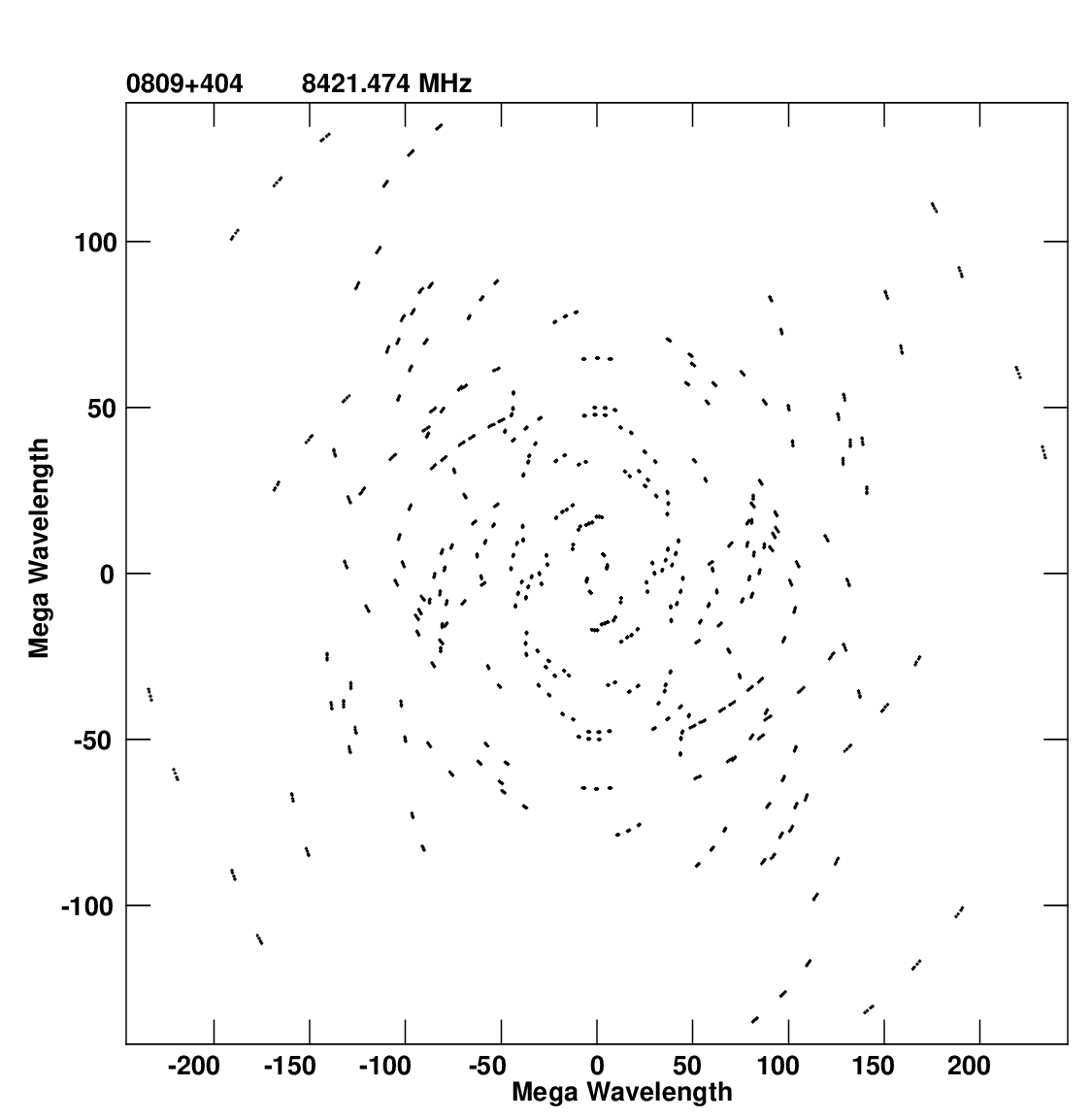}
\includegraphics[width=7cm, height=7cm]{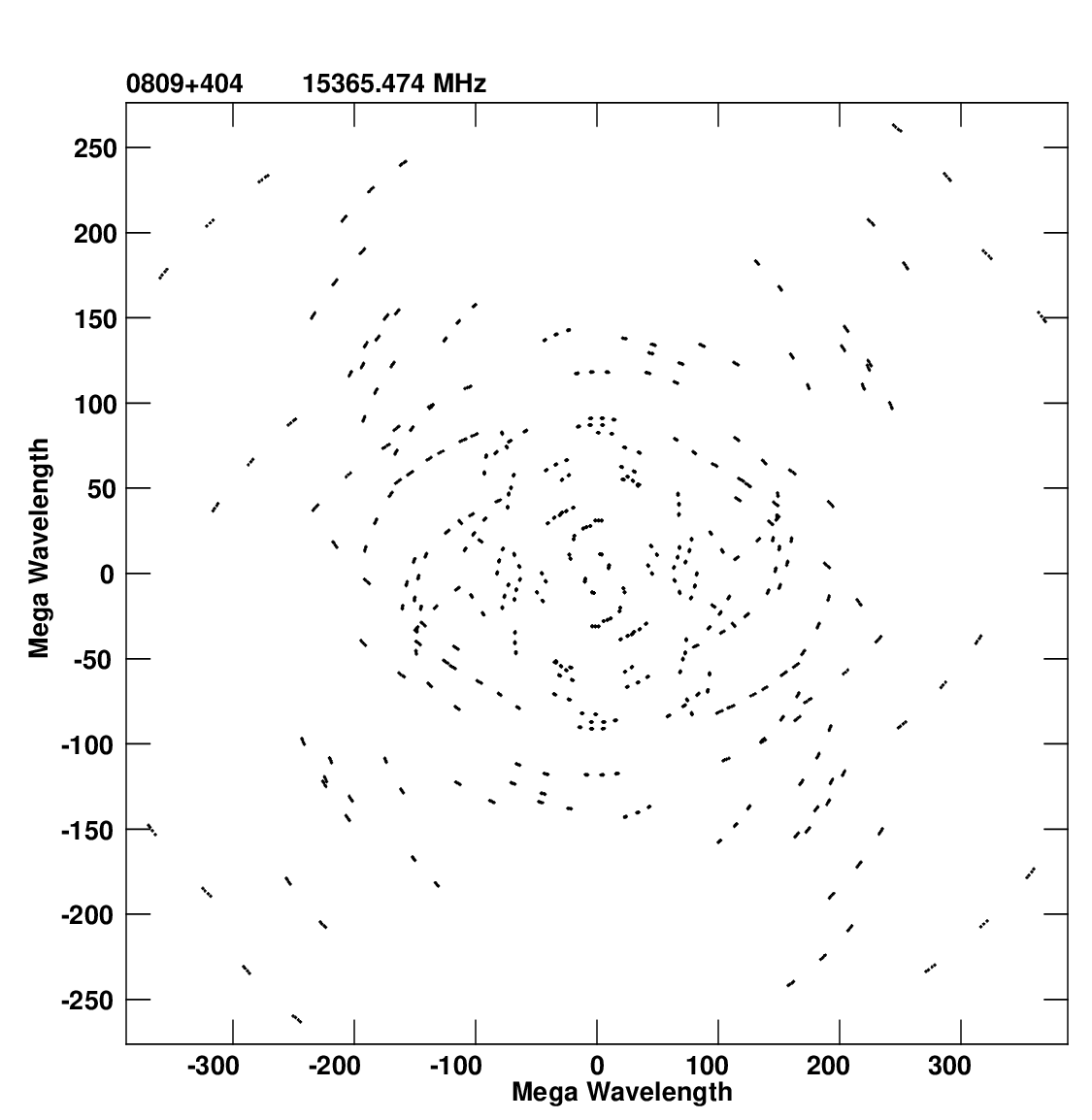}
\caption{Typical VLBA $u$-$v$ coverage at four different frequencies
attained during the observations of 0809+404.}
\label{uv_coverage}
\end{figure*}

The $u$-$v$ coverage for the observations  of the source 0809+404 at all
four frequencies, that are typical for all the observations, are shown in
Fig.~\ref{uv_coverage}.

The whole data reduction process was carried out using AIPS. The data
obtained during the second observational campaign (at 5, 8.4 and 15.4\,GHz)
were affected by Earth Orientation Parameter (EOP) errors introduced by the
VLBA correlator. These errors were first removed in the reduction process. 
Residual fringe delay and rate corrections derived for the appropriate
phase-reference sources, were applied to the corresponding target source data
and initial images of these were produced. For a majority of the target
sources, these images and their corrected data were then used as input
parameters for further cycles of phase self-calibration. In most cases
amplitude self-calibration was also applied. The final ``naturally weighted''
images were produced using IMAGR. The total intensity
images are shown in Figs.~\ref{0809+404_maps} to~\ref{1616+366_maps}. Flux
densities of the principal components of the sources were measured using the
AIPS task JMFIT and are listed in Table~\ref{table2}.

It was realised that, because of poor $u$-$v$ coverage and a ``break-up''
of the structure of some sources at the higher frequencies, appreciable amounts
of flux density could be missing. Therefore, spectral index maps that were
tentatively produced were not considered to be reliable and
it was decided not to include them, nor the integrated spectra in this paper.
Also the spectral indices one can calculate from the flux densities quoted
in Table~\ref{table2} should only be treated as a coarse approximation.

In addition to the observations described above, an unpublished 15-GHz VLA
observation of 0809+404, made in A-conf. by B.~Clark and R.~Perley in
November~1983 as a part of investigations of the B3-VLA sample \citep{vig89},
has been included here with the authors' kind permission
(Fig.~\ref{0809+404_maps}, upper left panel).

\section{Comments on individual sources} 

\begin{figure*}[t]
\centering
\includegraphics[width=8cm, height=8cm]{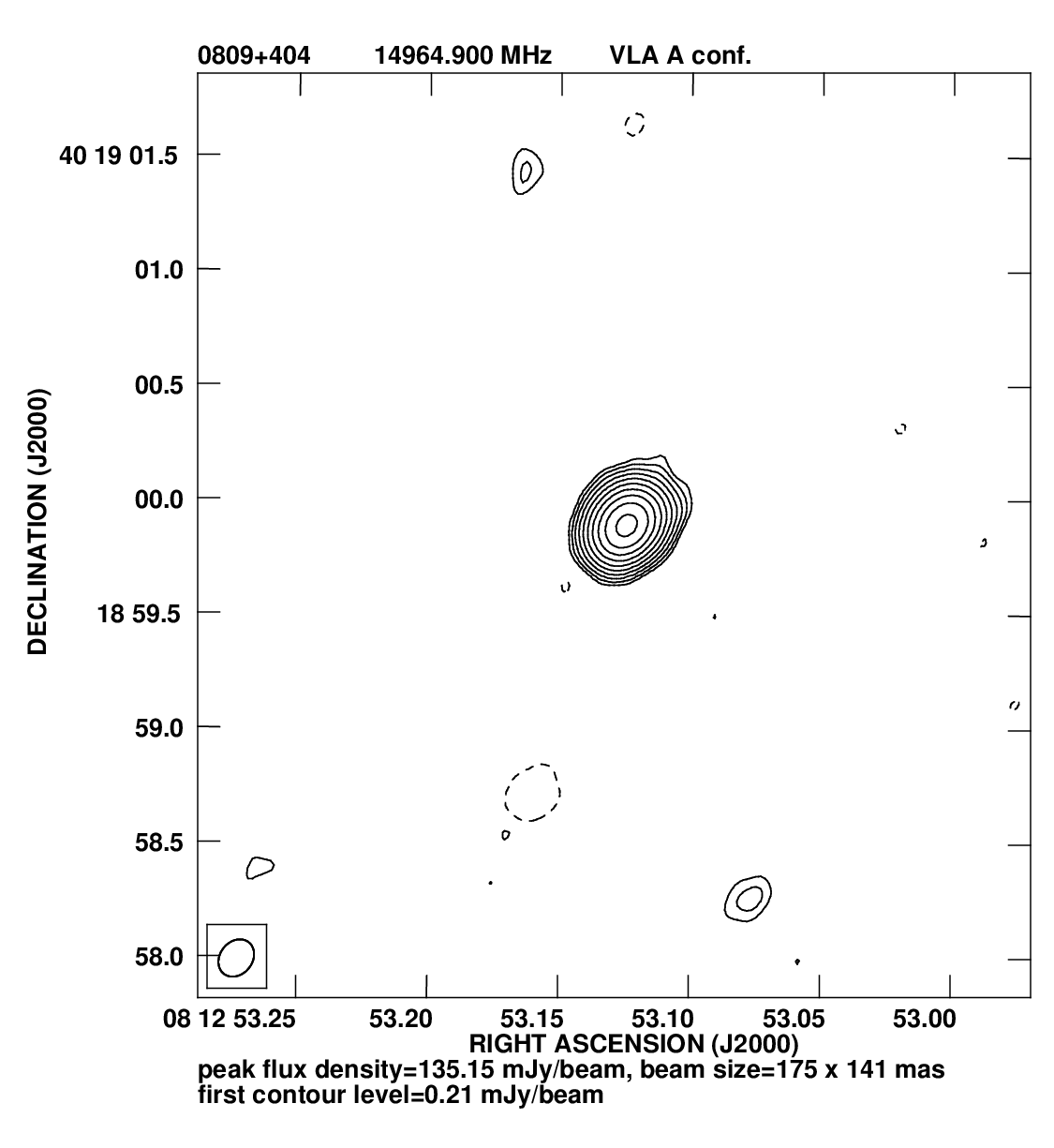} 
\includegraphics[width=8cm, height=8cm]{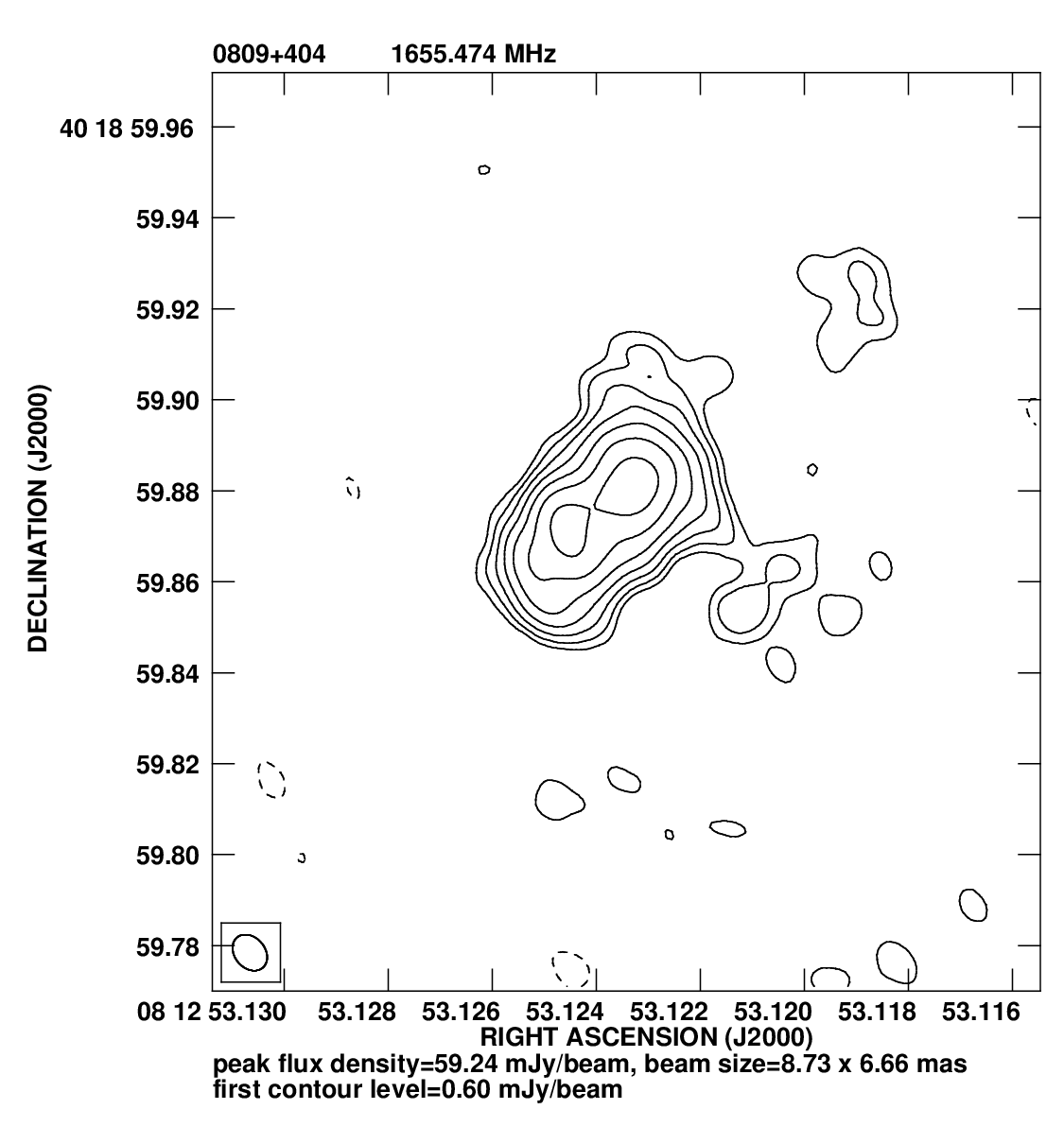}
\includegraphics[width=8cm, height=8cm]{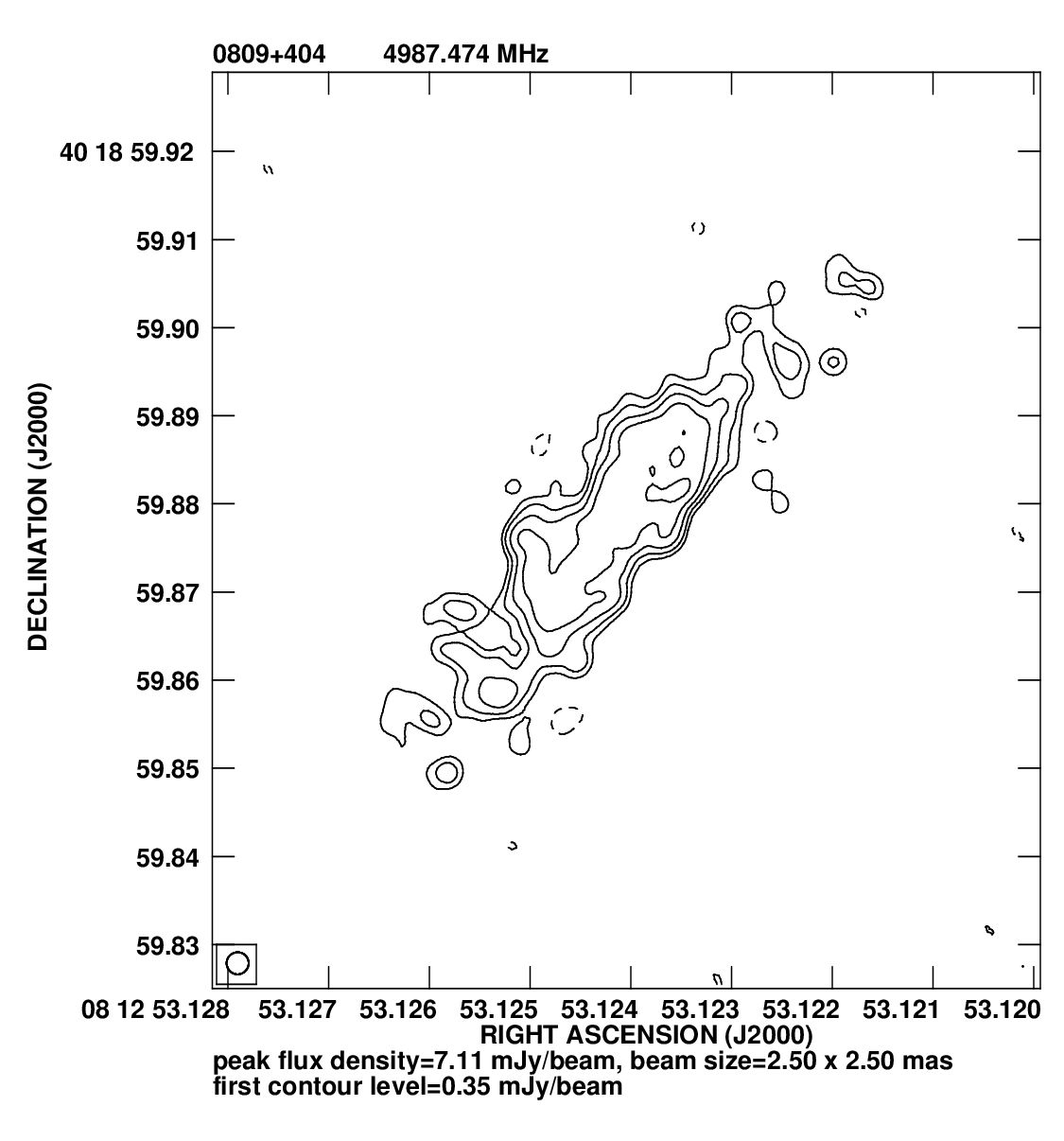}
\includegraphics[width=8cm, height=8cm]{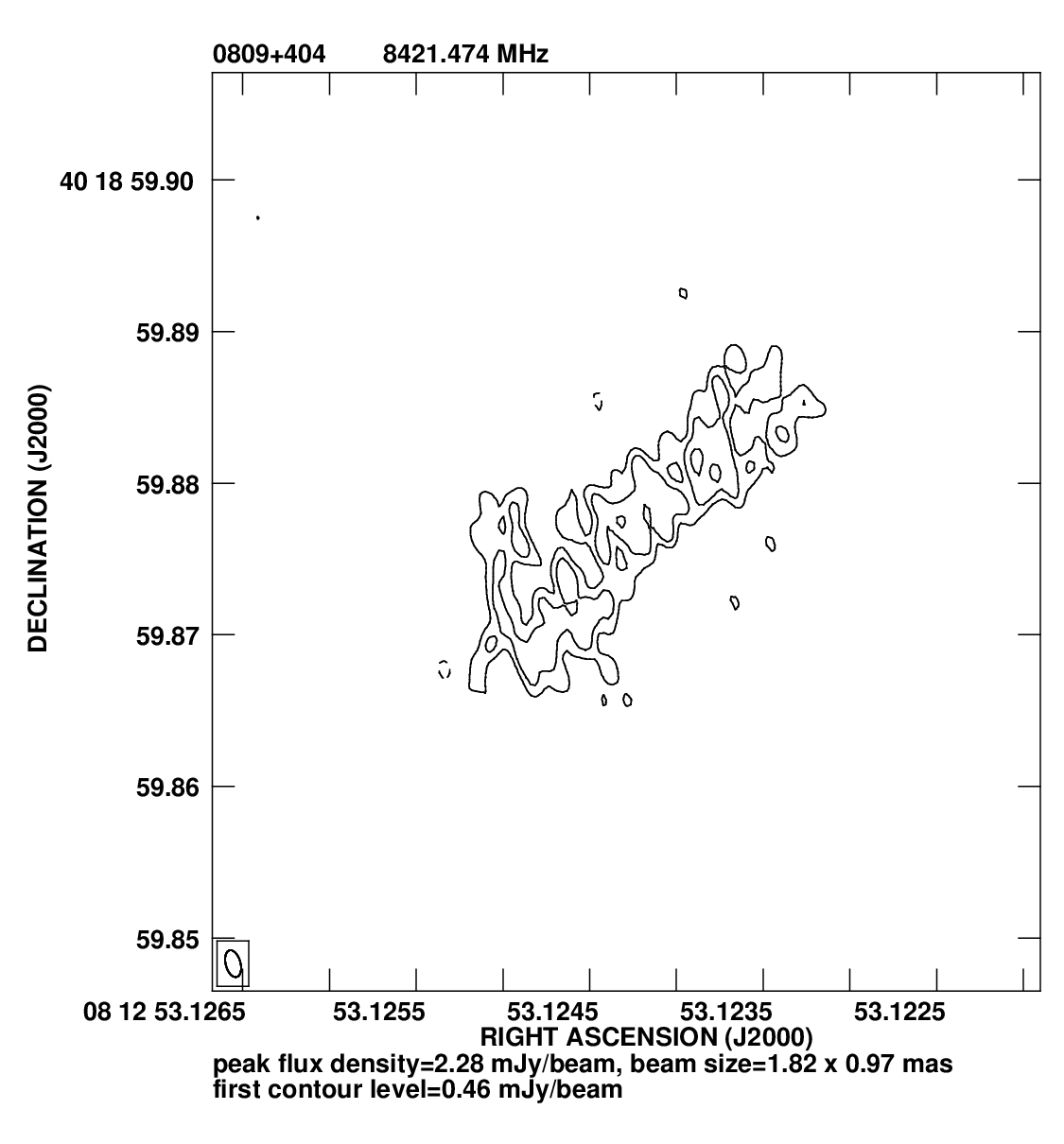}
\caption{The VLA (A-conf.) 15\,GHz map (upper left) and VLBA maps of 0809+404
at 1.65, 5 and 8.4\,GHz. Contours increase
by a factor 2 and the first contour level corresponds to $\approx 3\sigma$.}
\label{0809+404_maps}
\end{figure*}

\noindent {\bf \object{0809+404}}. VLA observations at 4.9 and 8.5\,GHz by
\citet{f2001} -- hereafter F2001 -- show the source to have a highly
asymmetric double structure with the two components separated by 1\farcs2
and with a flux density ratio of $\sim$100:1 at 4.9\,GHz. The much weaker
western component is somewhat resolved with the VLA at 8.5\,GHz and, not
surprisingly, is not visible in the VLBA image of \citet{dal02} -- hereafter
D2002 -- at 1.67\,GHz, nor in our VLBA image at approximately the same
frequency (Fig.~\ref{0809+404_maps}), that looks very similar to the image
of D2002. The eastern component, compact at VLA resolutions at the
lower frequencies, has been resolved with the VLBA at 1.67\,GHz into a
structure that could be a double. However, this may just be an artefact of
the contouring and D2002 consider it to have an amorphous structure.
Our VLBA images at 5\,GHz and 8.4\,GHz (Fig.~\ref{0809+404_maps}) and our
inability to detect the source in our 15.4-GHz VLBA observations, confirm the
latter. The source is simply fading away at the higher frequencies and there
is no indication of a core or hotspots. The tongue of emission pointing to
the south-west in the 1.65-GHz image is also seen in the image of
D2002, although the rather extended emission at its extremity in our
image, and which is absent in that of D2002, is probably an artefact
resulting from more extended emission that is present and being resolved.

The 15-GHz VLA (Fig.~\ref{0809+404_maps}) observations show only the brighter
eastern component which remains mostly unresolved and there is
no hint of a radio core in this image. The total flux density of this component
at 15\,GHz is 146.8 mJy, which yields a spectral index $\alpha=-1.03$ 
between 8.5 and 15\,GHz.

Based upon the value of the spectral index between
1.4 and 4.85\,GHz ($\alpha=-0.81$) and the total flux densities at those
frequencies (see Table~\ref{table1}), the interpolated total flux density
of the source at 1.65\,GHz is 932\,mJy. This means that $\sim$48\% of
the total flux density has been missed in our VLBA image which can be
partially explained by the fact that only the brighter component in the VLA
images made by F2001 has been detected. Moreover, compared with
the flux densities of the brighter component of the source at 4.9 and 8.5\,GHz
measured by F2001, it appears that only $\sim$47\% and $\sim$28\% of its flux
density has been detected at the two corresponding frequencies
in our VLBA observations.

According to \citet{vig97} 0809+404 is a Seyfert galaxy with redshift
$z=0.551$. A galaxy is also the (automated) morphological identification
of this object (RA=\,$8^{\rm h}12^{\rm m}53\fs109$,
Dec=\,$+40\degr 19\arcmin 0\farcs00$)
included in Data Release~4 of the Sloan Digital Sky Survey
(SDSS/DR4), the latest release of SDSS at the time of writing. The redshift
quoted by SDSS is in a full agreement with that of \citet{vig97}. However,
spectroscopically 0809+404 appears in SDSS as a quasar which is perhaps a more
appropriate identification given that 0809+404 is included in the list of
type-II quasar candidates \citep{zakam03} -- see further discussion in
Subsection~\ref{s-0809}.

\begin{figure*}
\centering
\includegraphics[width=5.5cm, height=11cm, angle=-90]{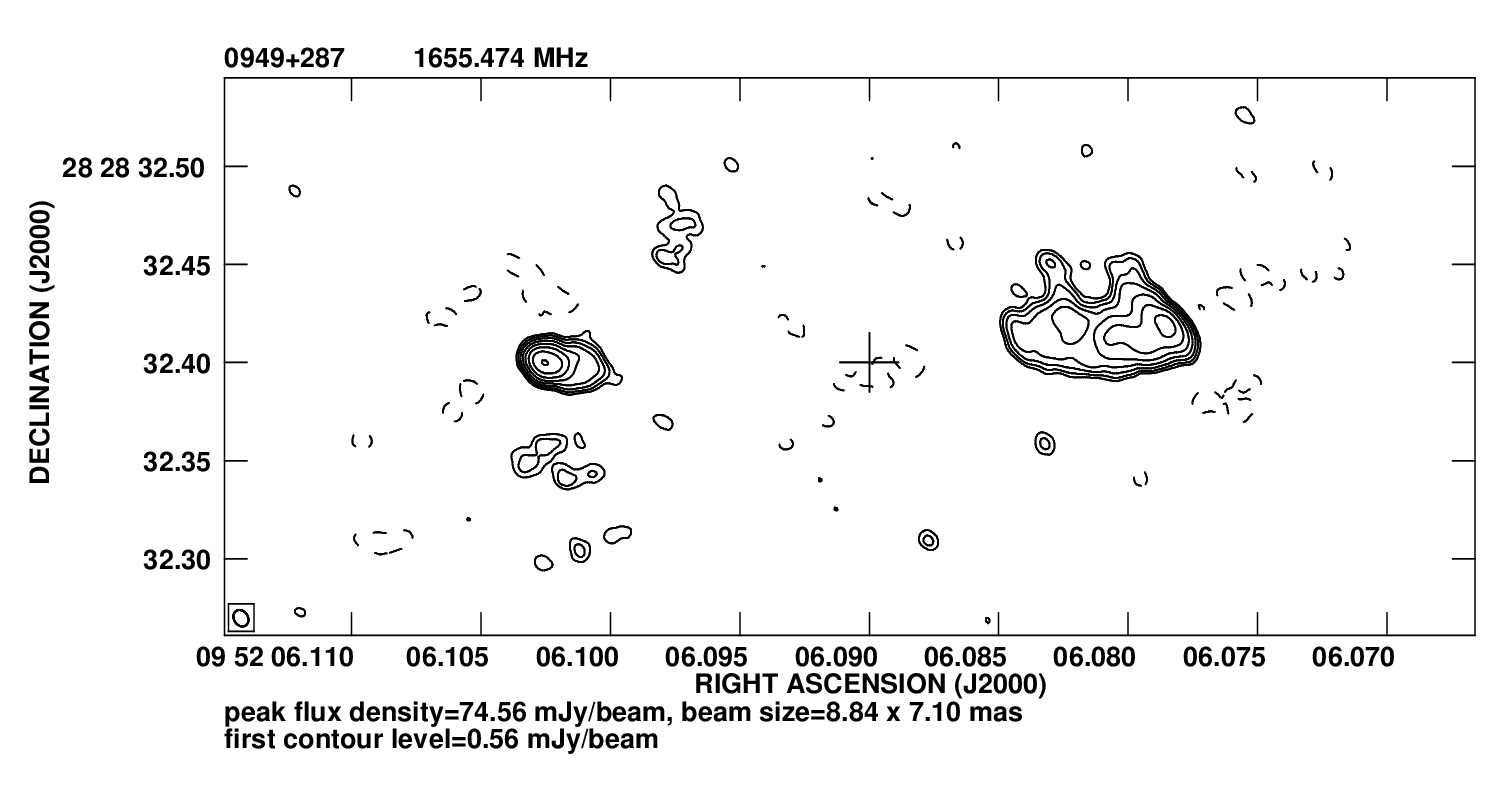}
\begin{center}
\includegraphics[width=5.5cm, height=11cm, angle=-90]{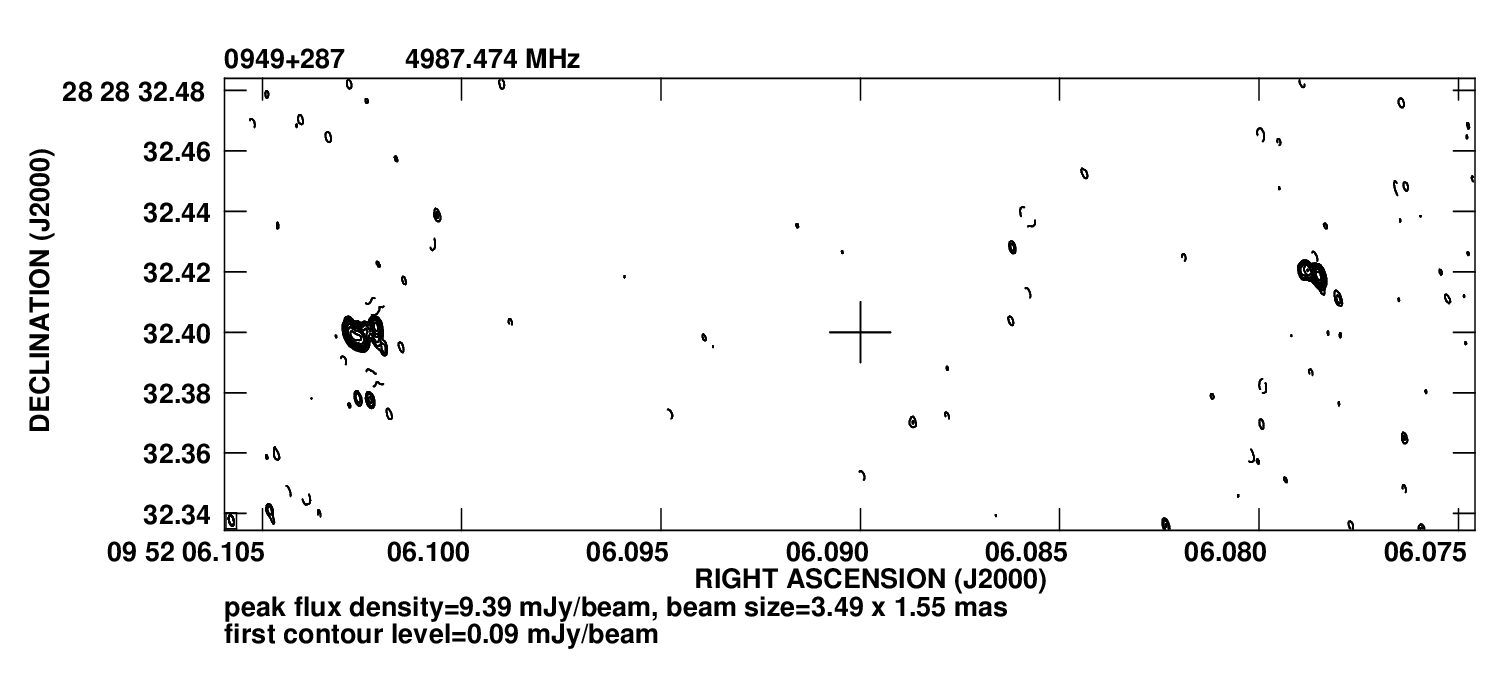}
\end{center}
\includegraphics[width=6cm, height=6cm]{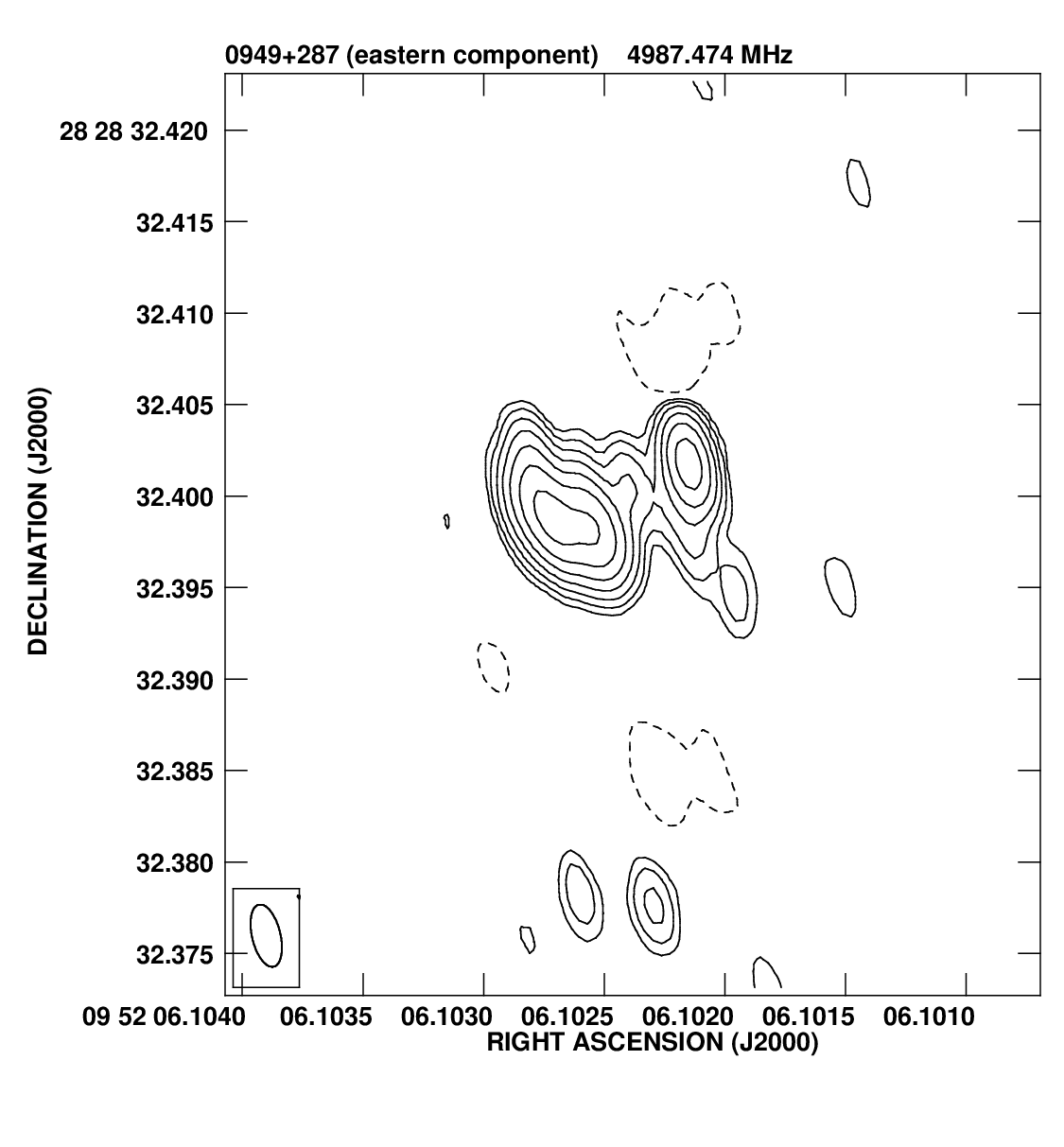}
\hspace{1cm}
\includegraphics[width=6cm, height=6cm]{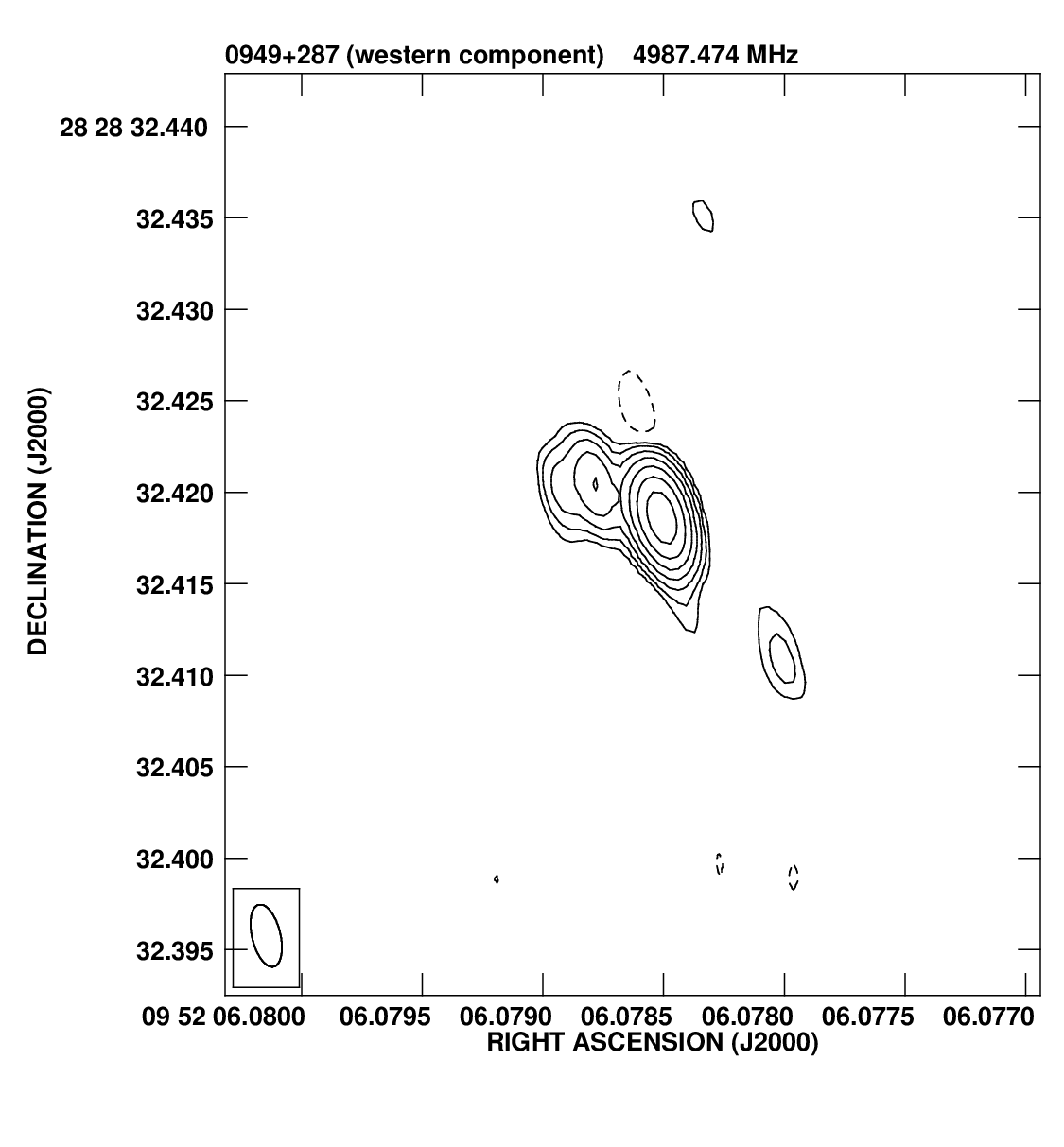}
\caption{The VLBA maps of 0949+287 at 1.65\,GHz (top) and 5\,GHz (middle). 
The bottom row of panels show enlargements of the main components
of the source at 5\,GHz. Contours increase
by a factor 2 and the first contour level corresponds to $\approx 3\sigma$.
The position of the optical object extracted from SDSS/DR4 is marked with a
cross.}
\label{0949+287_maps}
\end{figure*}

\noindent {\bf \object{0949+287}}. Our VLBA radio images show a double structure
for this source (Fig.~\ref{0949+287_maps}),
the two detected components appearing to be radio lobes with steep spectra
between 1.65 and 8.4\,GHz. There is a compact feature in the eastern radio
lobe, that is becoming weaker at the higher frequencies.
The more diffuse western lobe also contains a small compact component
with a spectrum which is becoming slightly flatter
towards higher frequencies and so is most probably a hotspot. 
The two features on the
northern edge of the western lobe in the 1.65\,GHz image are probably
artefacts. There is no indication of a core in any of the images.

The assumed total flux density at 1.65\,GHz is 1201\,mJy which indicates that
$\sim$44\% of the total flux density of the source has not been seen
because of resolution. The optical object extracted from
SDSS/DR4 is a galaxy and its position
(RA=\,$9^{\rm h}52^{\rm m}6\fs098$, Dec=\,$+28\degr 28\arcmin 32\farcs40$)
is marked with a cross on all maps. It is to be noted that there is
another galaxy with $m_R$=21.74 located 2\farcs5 south of the galaxy
identified with the radio source.

The double structure of 0949+287 has been recently confirmed by
22\,GHz VLA observations made by \citet{bol04} and its integrated
spectrum shown there confirms it to be a compact steep spectrum source. The
polarisation properties of 0949+287 are not known.

\begin{figure*}
\centering
\begin{center}
\includegraphics[width=5.5cm, height=11cm, angle=-90]{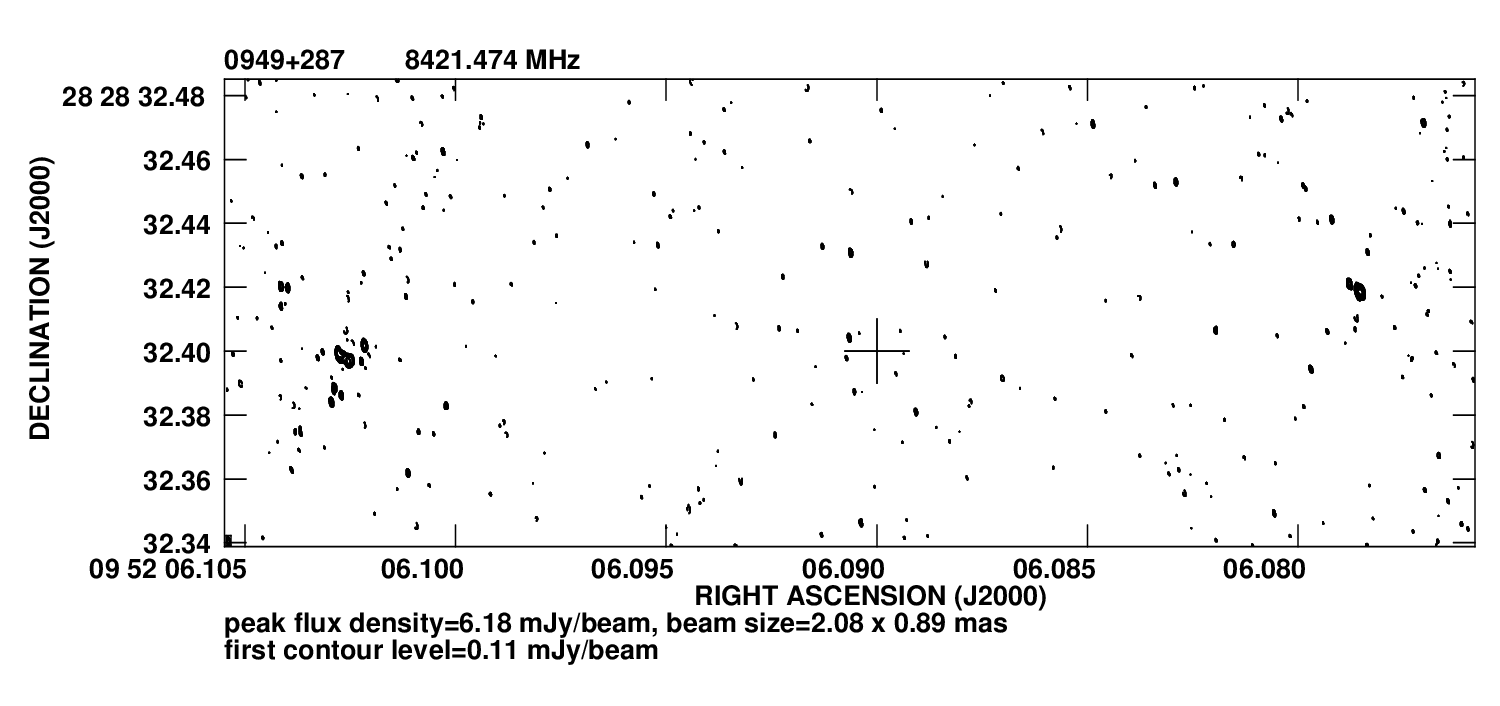}
\end{center}
\includegraphics[width=6cm, height=6cm]{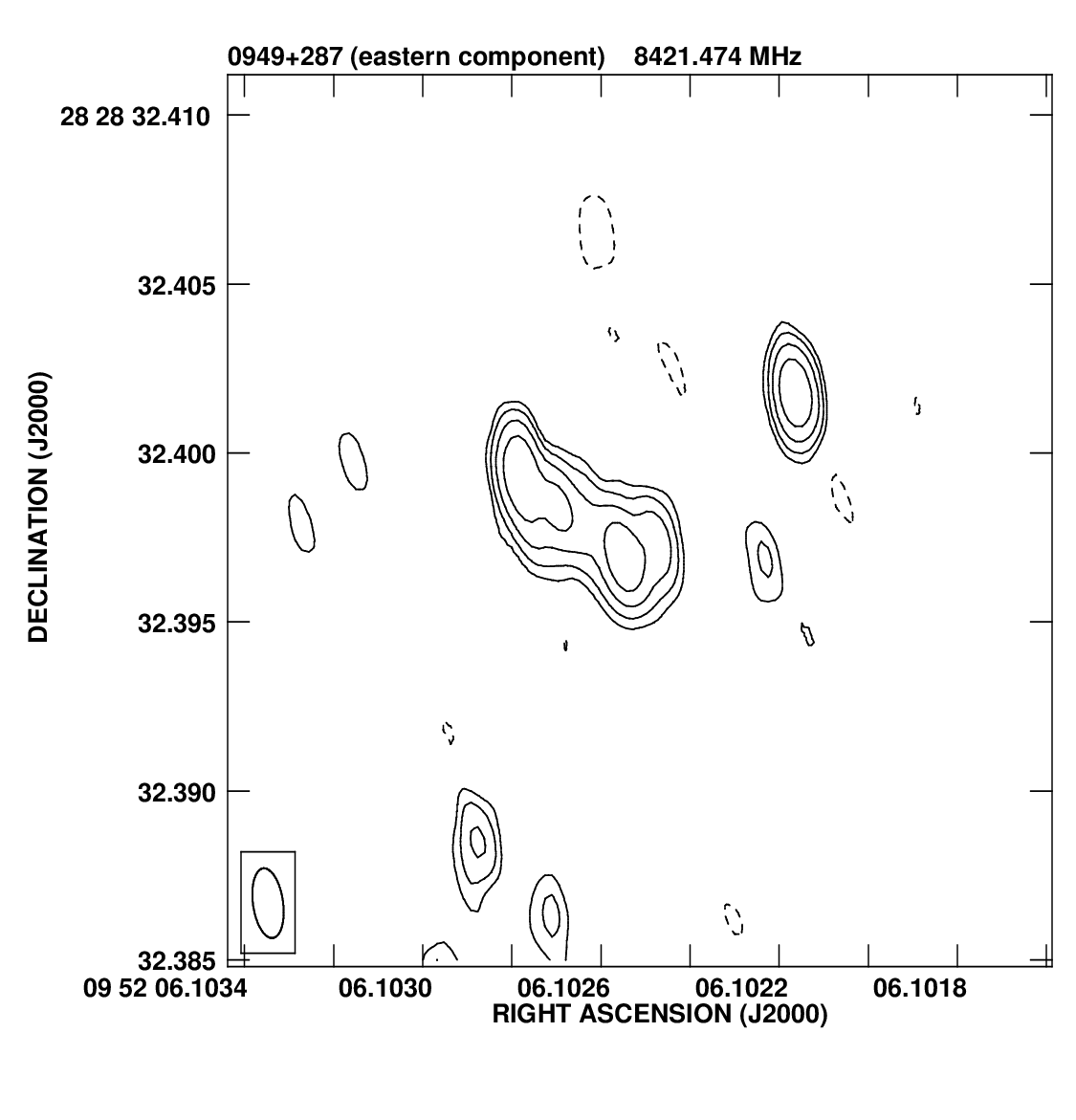}
\hspace{1cm}
\includegraphics[width=6cm, height=6cm]{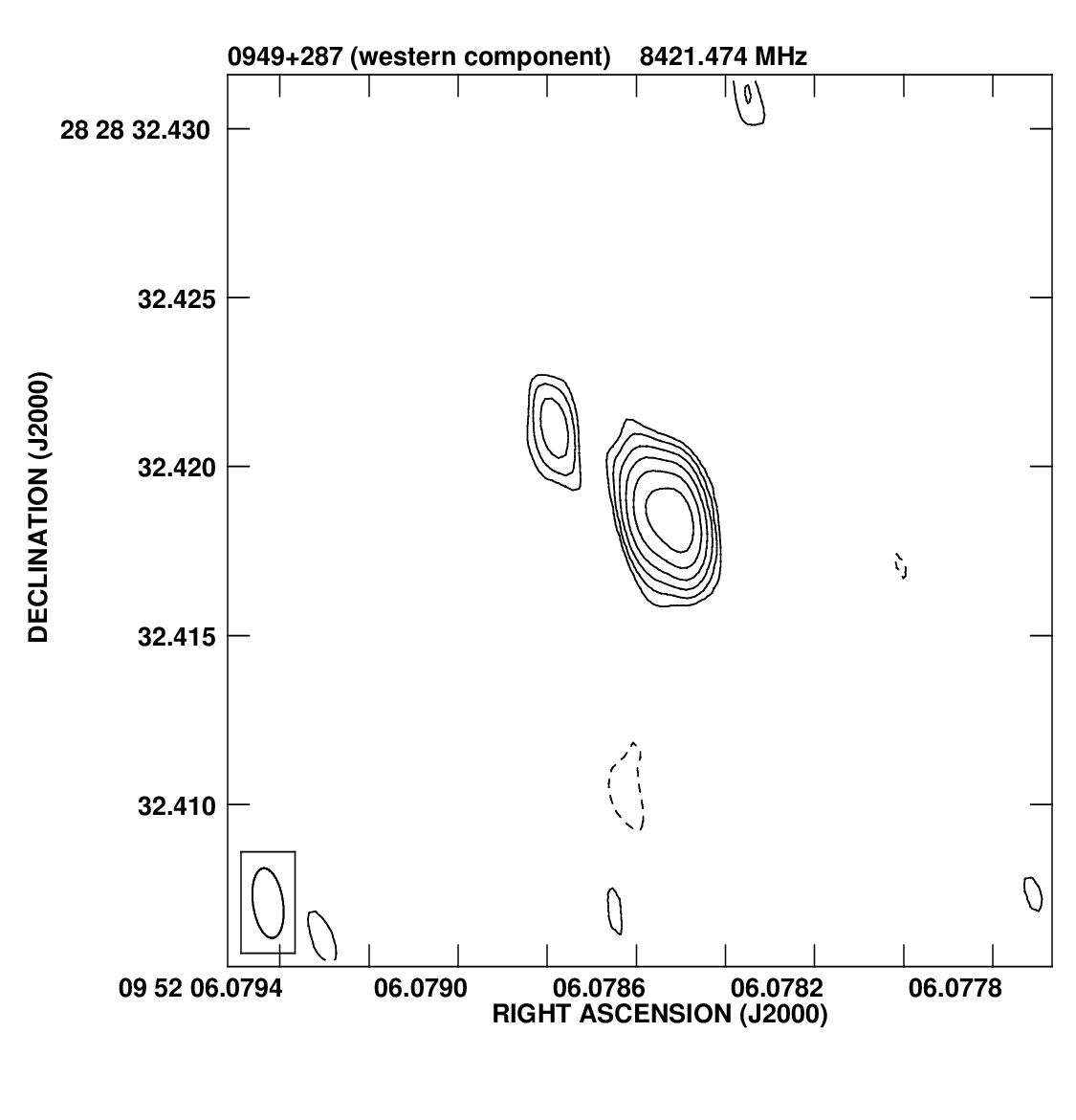}
\begin{flushleft}
{\bf Fig. \thefigure{}.} continued.
The VLBA map of 0949+287 (top) and its enlarged components (bottom) at
8.4\,GHz. 
Contours increase by a factor 2 and the first contour level corresponds to
$\approx 3\sigma$. The position of the optical object extracted from
SDSS/DR4 is marked with a cross.
\end{flushleft}
\end{figure*}

\begin{figure*}
\centering
\includegraphics[width=7cm, height=8cm]{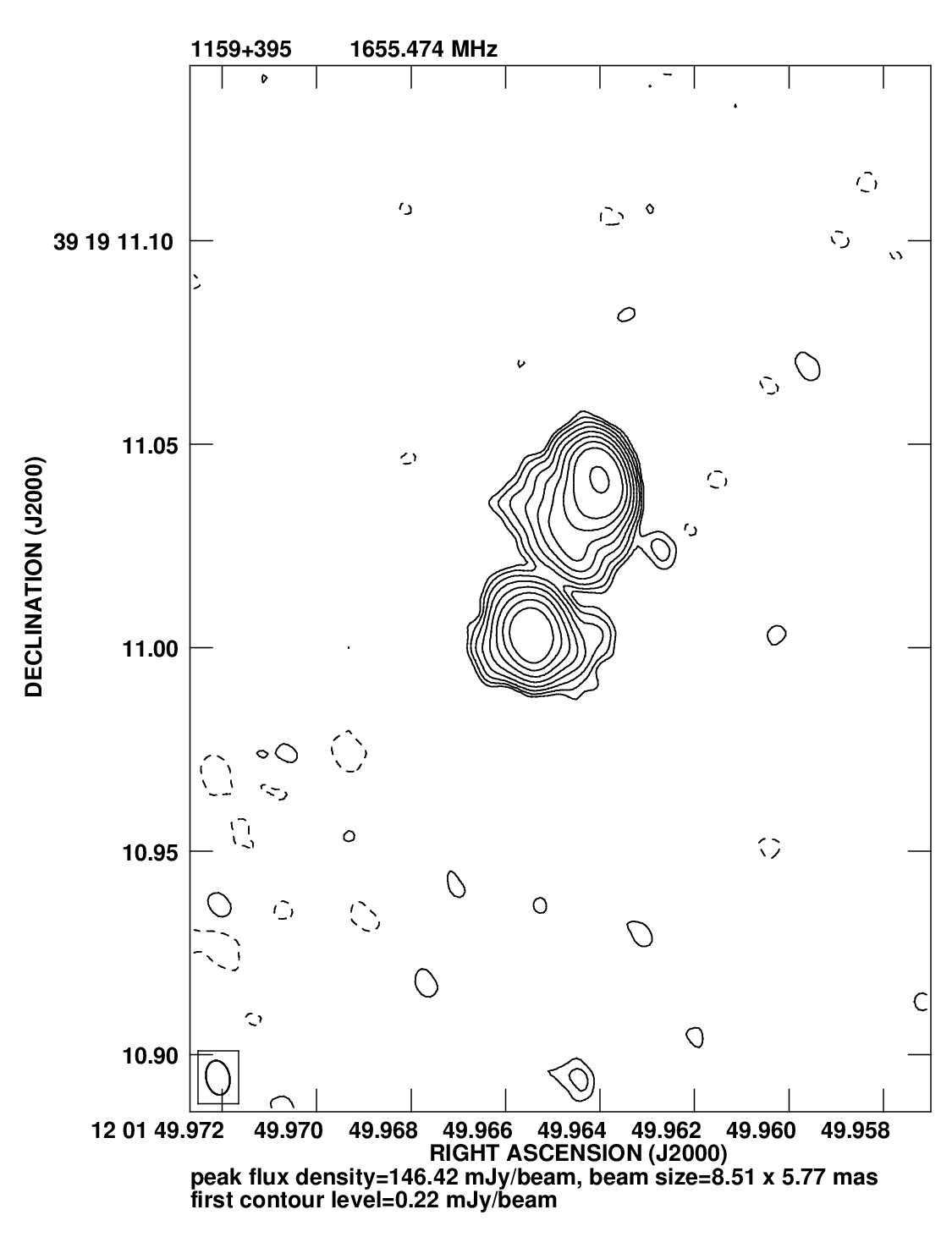}
\includegraphics[width=7cm, height=8cm]{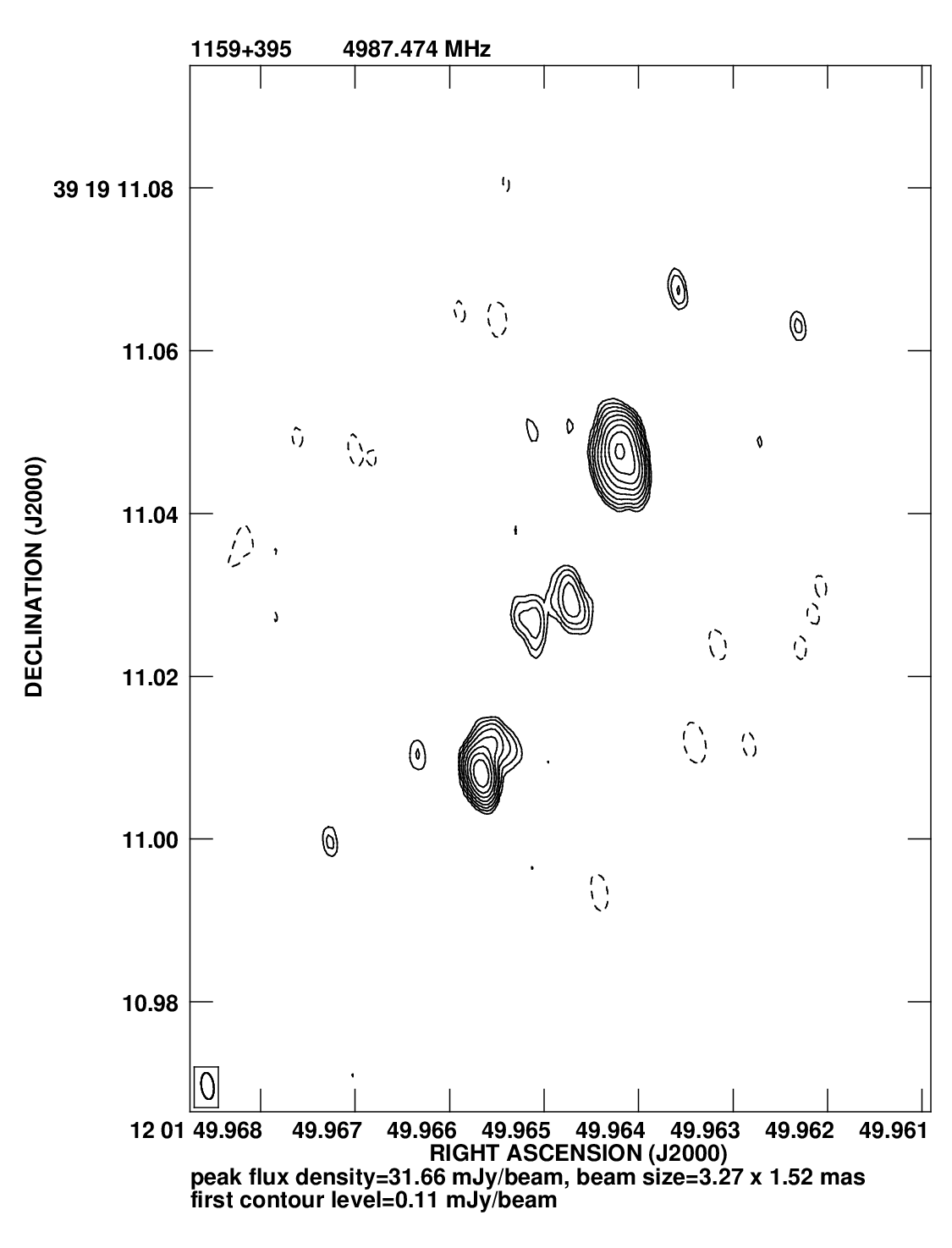}
\includegraphics[width=7cm, height=8cm]{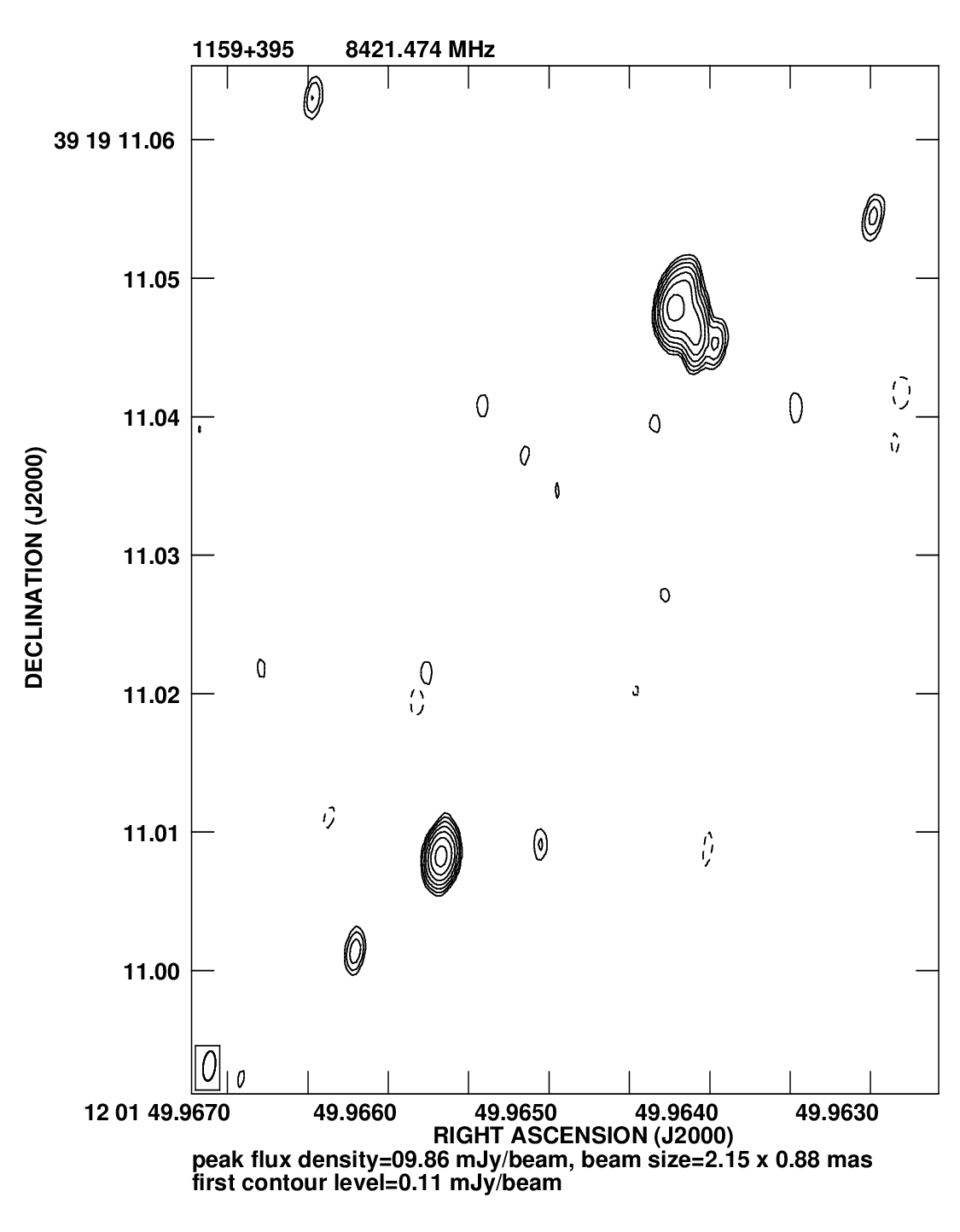}
\caption{The VLBA maps of 1159+395 at 1.65, 5 and 8.4\,GHz.
Contours increase by a factor 2 and the first contour level corresponds
to $\approx 3\sigma$.}
\label{1159+395_maps}
\end{figure*}

\noindent {\bf \object{1159+395}}. This source has been identified as a galaxy
with a redshift $z=2.37$ (F2001). 
A galaxy is also the (automated) morphological identification
of this object included in SDSS/DR4 (RA=\,$12^{\rm h}01^{\rm m}50\fs002$,
Dec=\,$+39\degr 19\arcmin 10\farcs95$), although the redshift is not given.
Our 1.65, 5 and 8.4-GHz images show 
that the source has a double structure, the two components of which are clearly
radio lobes oriented in a north-south direction (Fig.~\ref{1159+395_maps}).
The spectral
indices of the lobes between 1.65 and 5\,GHz and between 5 and 8.4\,GHz are
very steep. 
The elongated structure aligned east-west in the northern lobe in the 
1.65-GHz image is probably an artefact. The only indication of a core in any
of the images is in the 5-GHz image, in which there is a peak of emission at
RA=\,$12^{\rm h}01^{\rm m}49\fs965$, Dec=\,$+39\degr 19\arcmin 11\farcs028$.
However, this is very doubtful as there is no indication of a core at this
position in the other images. The double structure of the source has been 
confirmed in 1.65-GHz (D2002) and 5-GHz \citep{or04} VLBA images and
it has been classified as a CSO. The interpolated total flux density at 1.65\,GHz
is 533\,mJy, so our VLBA image accounts for $\sim$81\% of the total flux
density.

\begin{figure*}
\centering
\includegraphics[width=8cm, height=8cm]{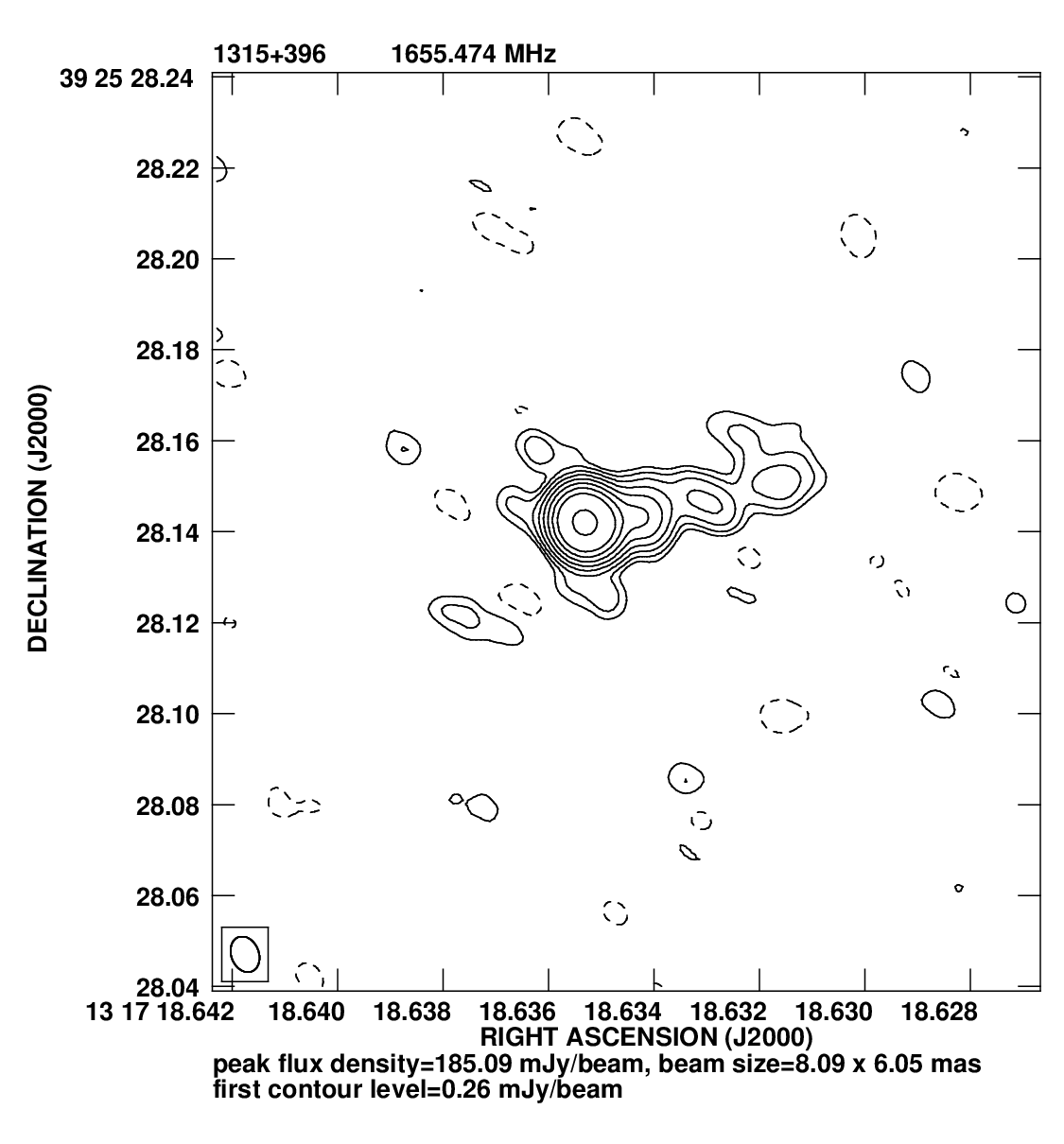}
\includegraphics[width=8cm, height=8cm]{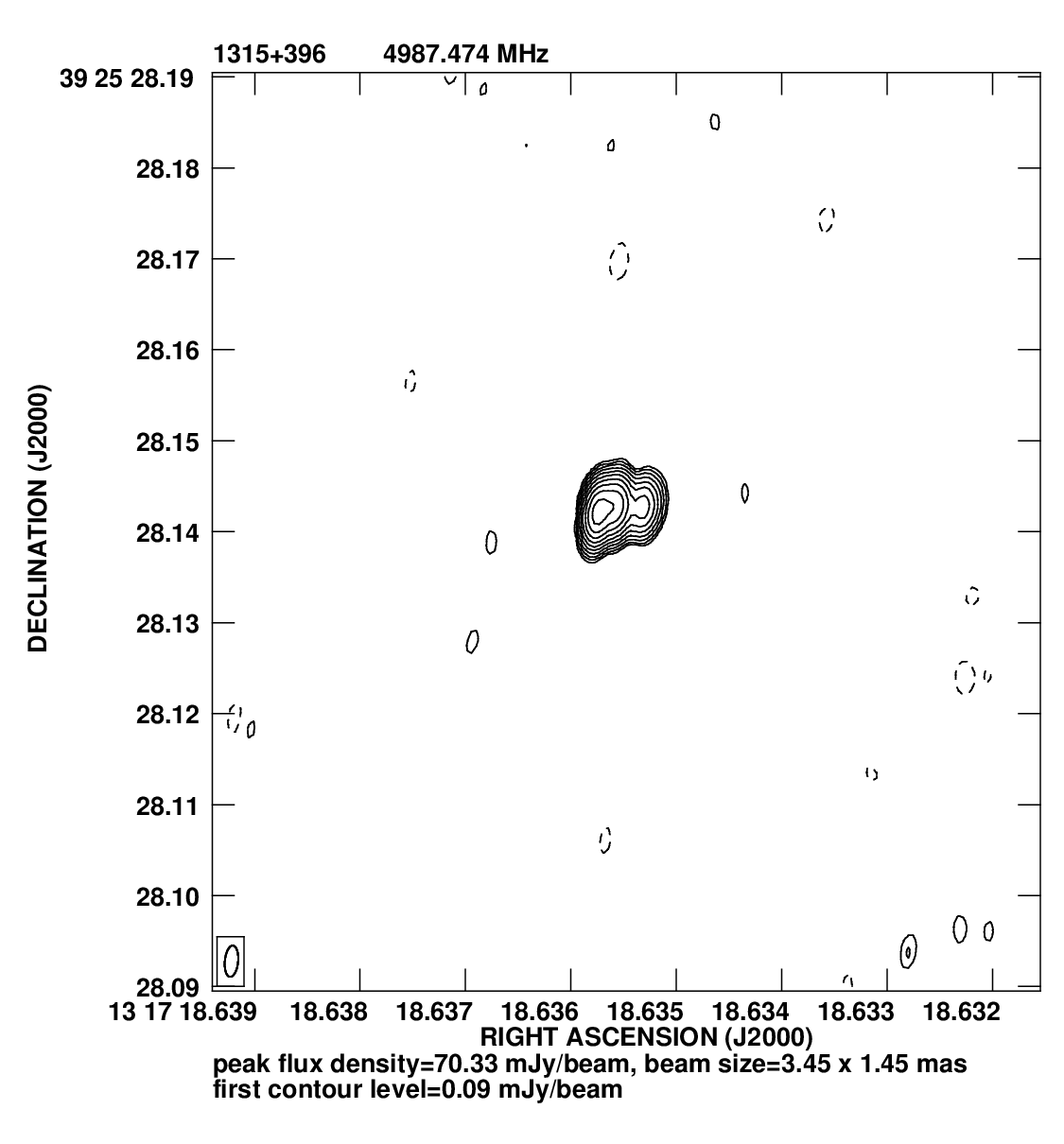}
\includegraphics[width=8cm, height=8cm]{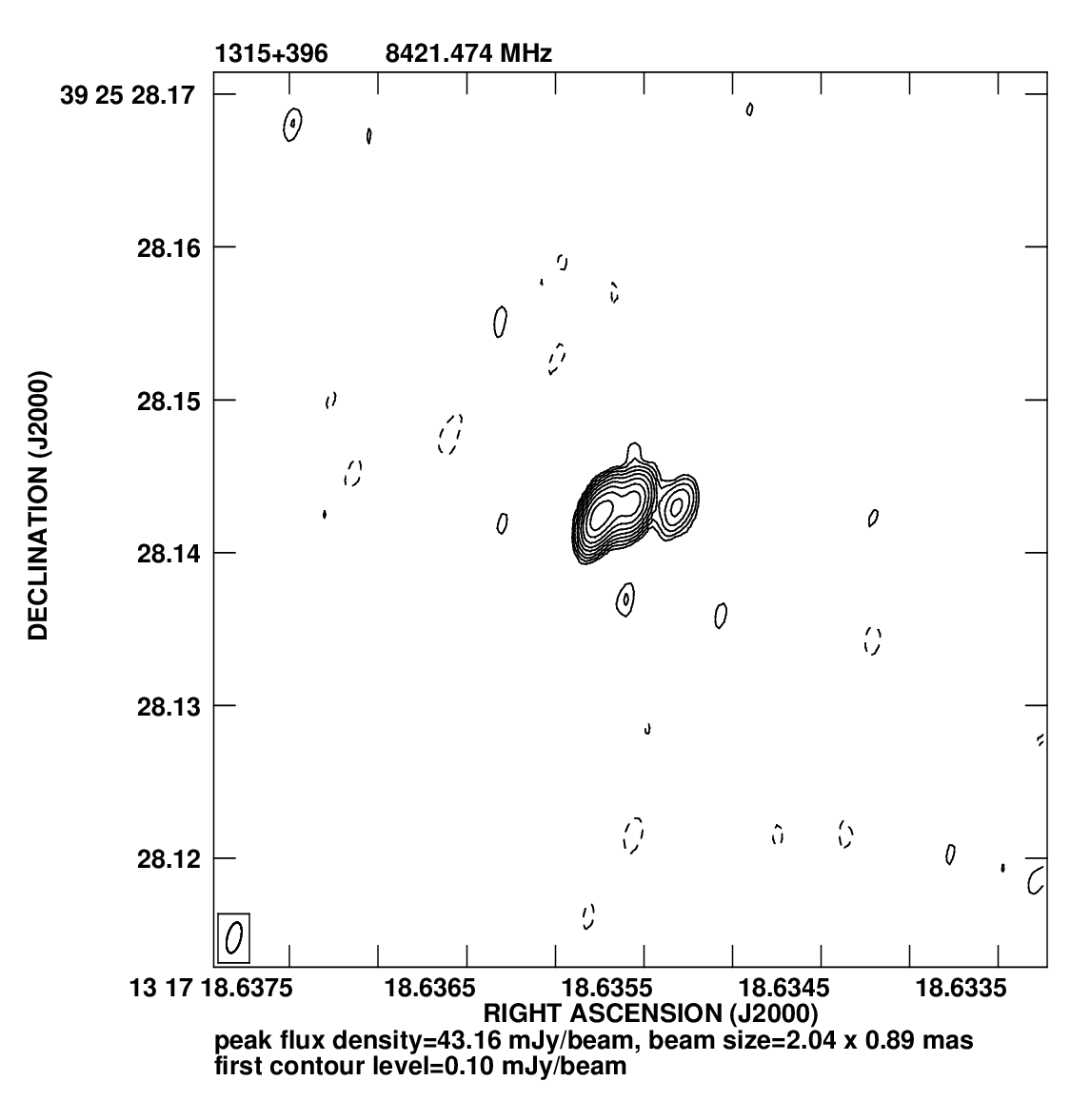}
\includegraphics[width=8cm, height=8cm]{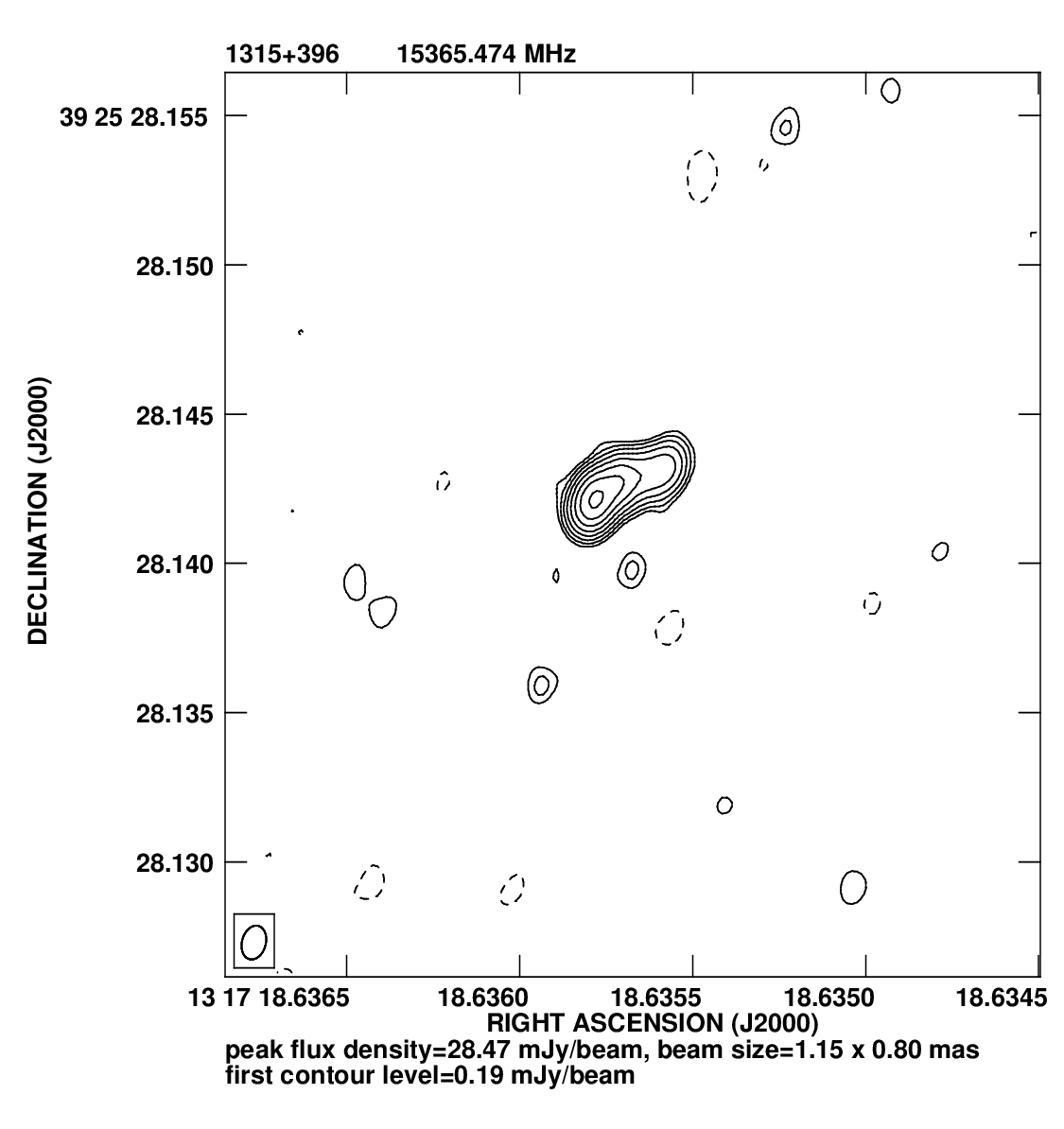}
\caption{The VLBA maps of 1315+396 at 1.65, 5, 8.4 and 15.4\,GHz.
Contours increase by a factor 2 and the first contour level corresponds
to $\approx 3\sigma$.}
\label{1315+396_maps}
\end{figure*}

\noindent {\bf \object{1315+396}}. The VLBA images indicate this to be a
core-jet source (Fig.~\ref{1315+396_maps}) that has been detected at all four
frequencies. The spectrum of the core component is steepening
towards higher frequencies. The 1.65-GHz image accounts
for $\sim$46\% of the total flux density of the source (538\,mJy), the latter
being derived from the flux densities and spectral index between 1.4 and
4.85\,GHz. The diffuse jet structure, visible in the 1.65-GHz image, could be
partially resolved by the VLBA. The source has been optically identified with a
quasar with a redshift $z=1.56$ \citep{vig90}. 
According to SDSS/DR4, 1315+396 is a star-like object with unknown redshift
at the position: RA=\,$13^{\rm h}17^{\rm m}18\fs643$,
Dec=\,$+39\degr 25\arcmin 28\farcs16$.

\begin{figure*}
\centering
\includegraphics[width=8cm, height=8cm]{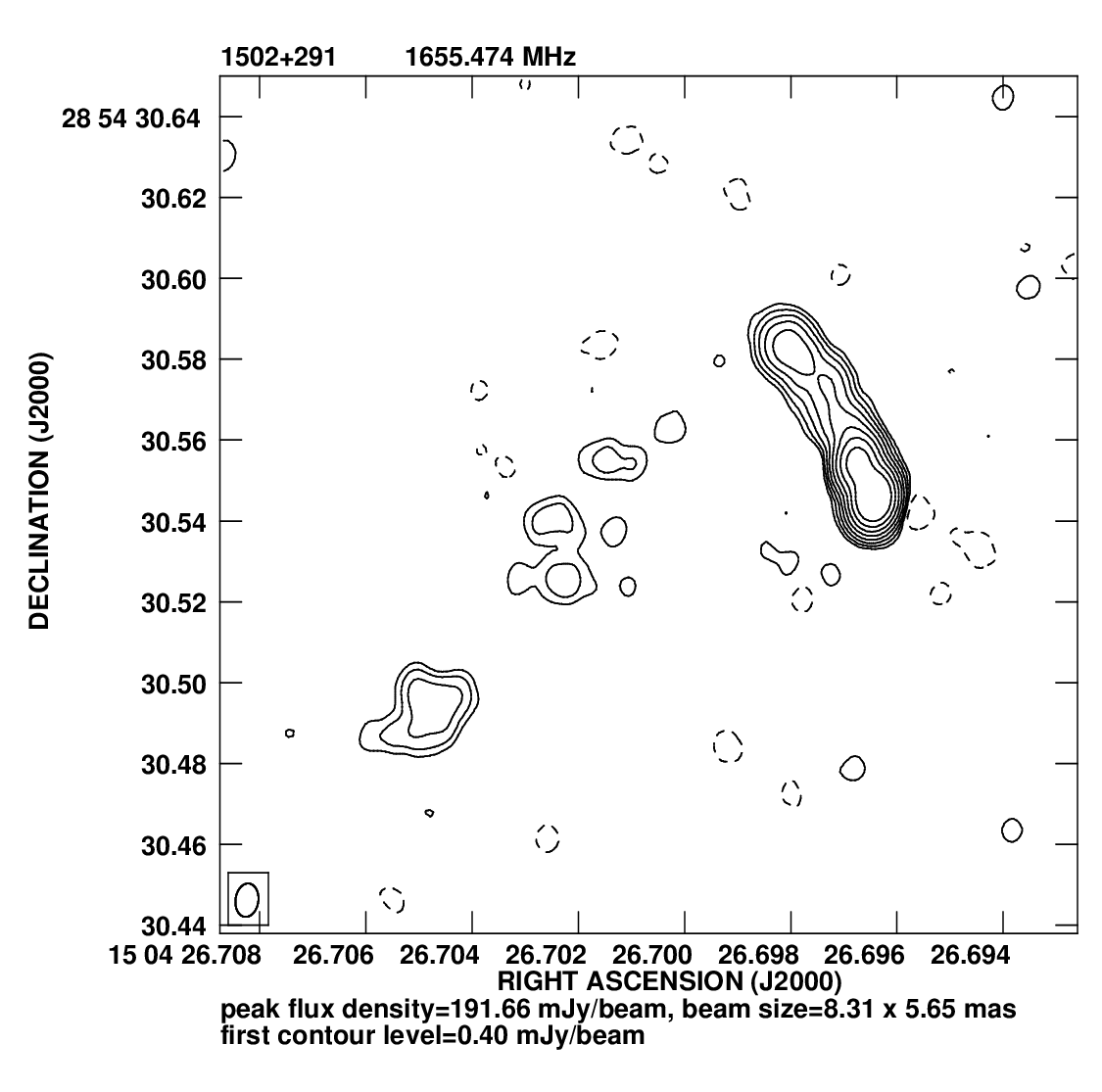}
\includegraphics[width=8cm, height=8cm]{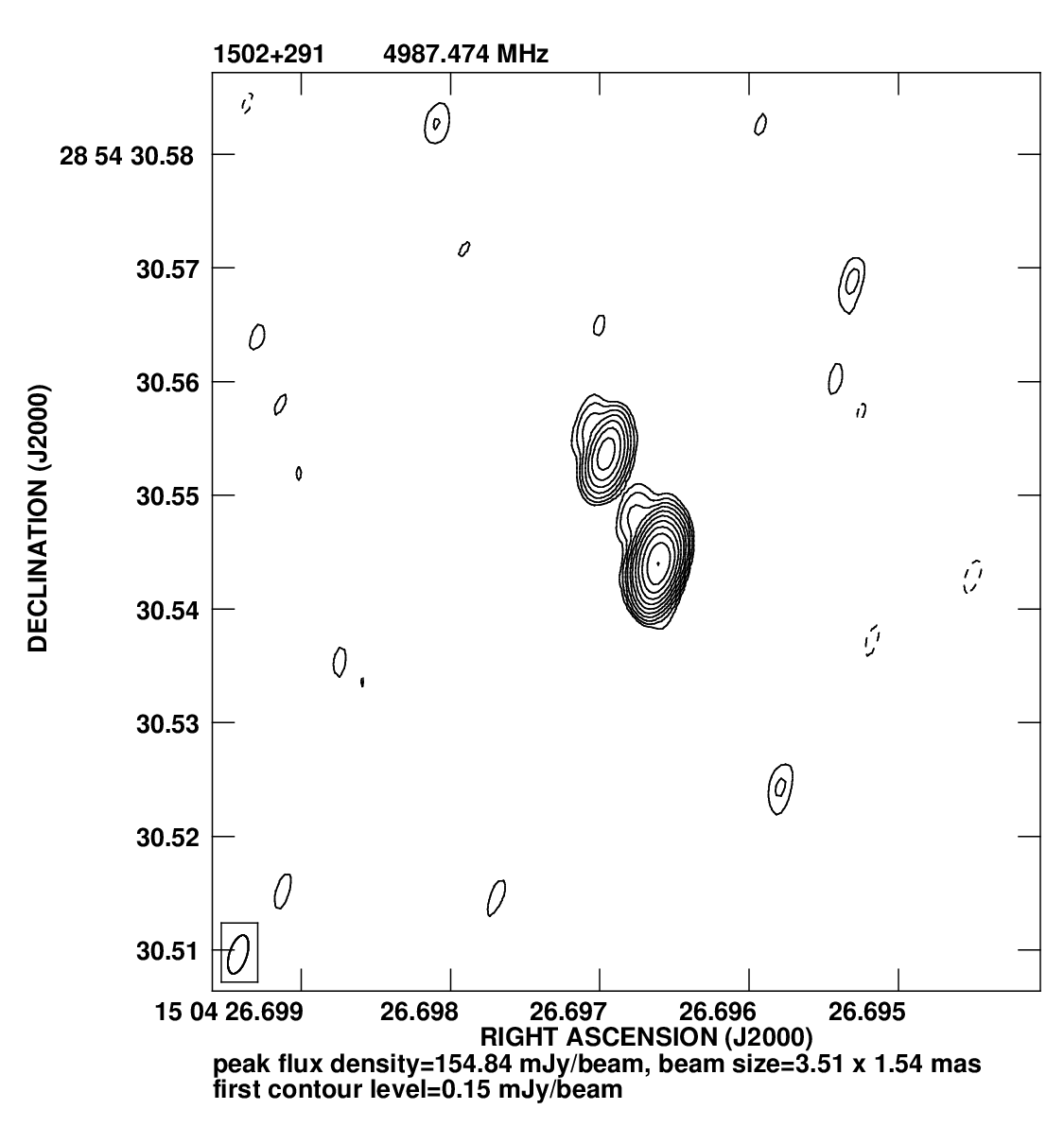}
\caption{The VLBA maps of 1502+291 at 1.65 and 5\,GHz.
Contours increase by a factor 2 and the first contour level corresponds
to $\approx 3\sigma$.}
\label{1502+291_maps}
\end{figure*}

\noindent {\bf \object{1502+291}}. The VLBA images of this source primarily
show an asymmetric structure directed to the north-east
(Fig.~\ref{1502+291_maps}).
As this structure was already clearly visible in the 1.65-GHz image
prior to the higher frequency observations, the source had already been
rejected as a candidate for a dying CSO and consequently was
only observed at 5\,GHz
to determine its spectral index and to confirm its core-jet nature. The peak
of emission at RA=\,$15^{\rm h}04^{\rm m}26\fs696$,
Dec=\,$+28\degr 54\arcmin 30\farcs545$ (Table~\ref{table2}) has a flat spectrum
and is a radio core. Approximately 70\% of the total flux density
(510\,mJy) of 1502+291 has been seen in our 1.65-GHz image.
VLBA snapshot observations of 1502+291 at 2.3 and 8.4\,GHz were also
carried out as a part of the VLBA Calibrator Survey \citep[VCS1,][]{beas02}.
The 2.3\,GHz image shows a core-jet structure whereas at 8.4\,GHz only the core
is visible.

Of interest in the 1.65-GHz image of 1502+291 are two 
regions of diffuse emission located to the south-east of the core-jet object.
It could be that they are lobes of a very nearby (with respect to the ``main''
core-jet structure), relic, compact double system.
The flux density of the better defined, more southerly lobe has been measured 
(Table~\ref{table2}). 
These putative lobes of a dead source, are invisible in the 5-GHz image, so they
must have steep spectra. Full-track VLBI observations are necessary to confirm
the conjecture that there are two double sources in the field of 1502+291.
1502+291 is a member of cluster of galaxies (Abell
2022) and, based on this, an estimate of its redshift
($z=0.056$) has been made \citep{abell}.
According to SDSS/DR4, 1502+291 is a star-like object with unknown redshift
at the position: RA=\,$15^{\rm h}04^{\rm m}26\fs698$,
Dec=\,$+28\degr 54\arcmin 30\farcs54$.

\begin{figure*}
\centering
\includegraphics[width=8cm, height=8cm]{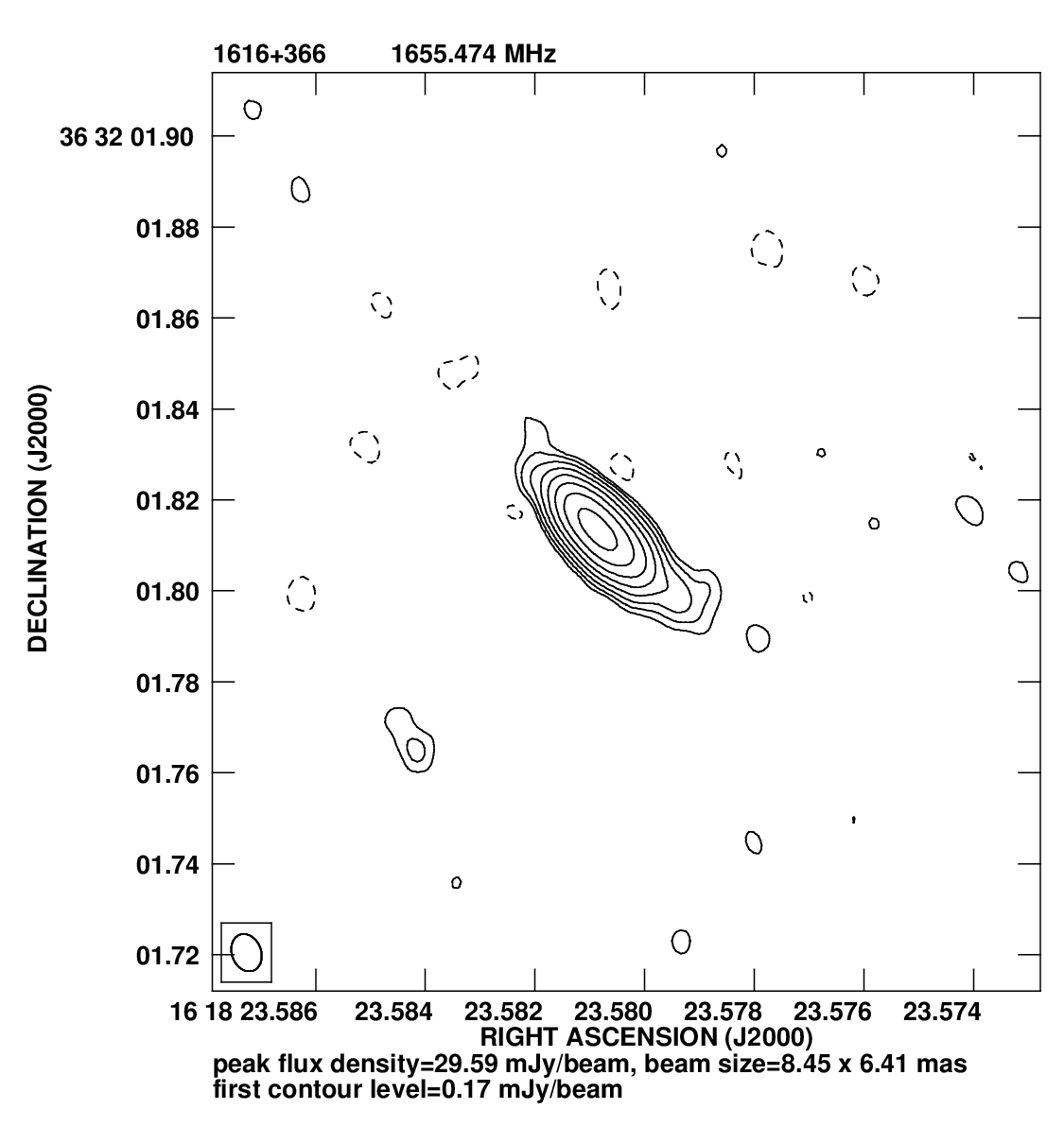}
\includegraphics[width=8cm, height=8cm]{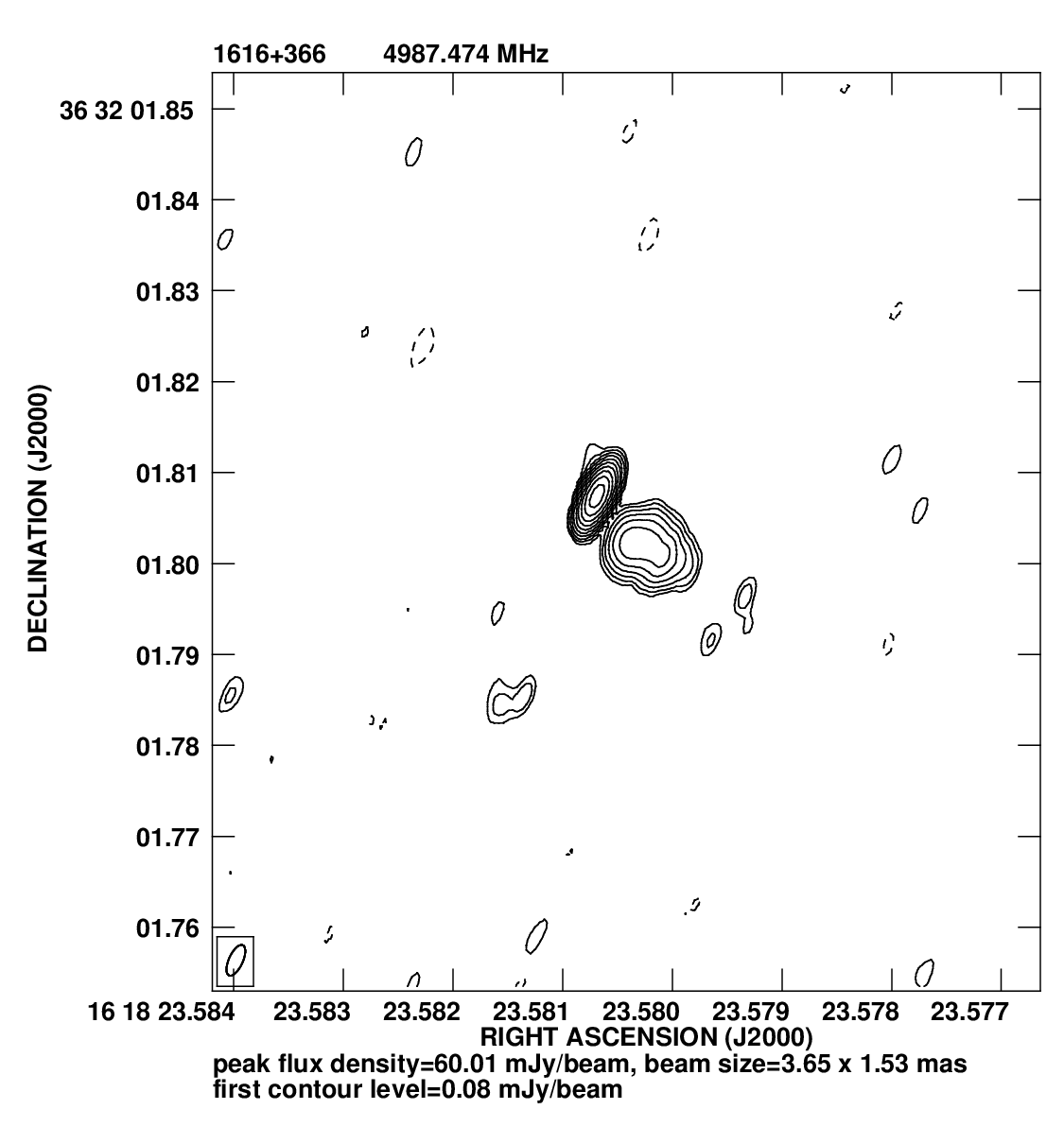}
\includegraphics[width=8cm, height=8cm]{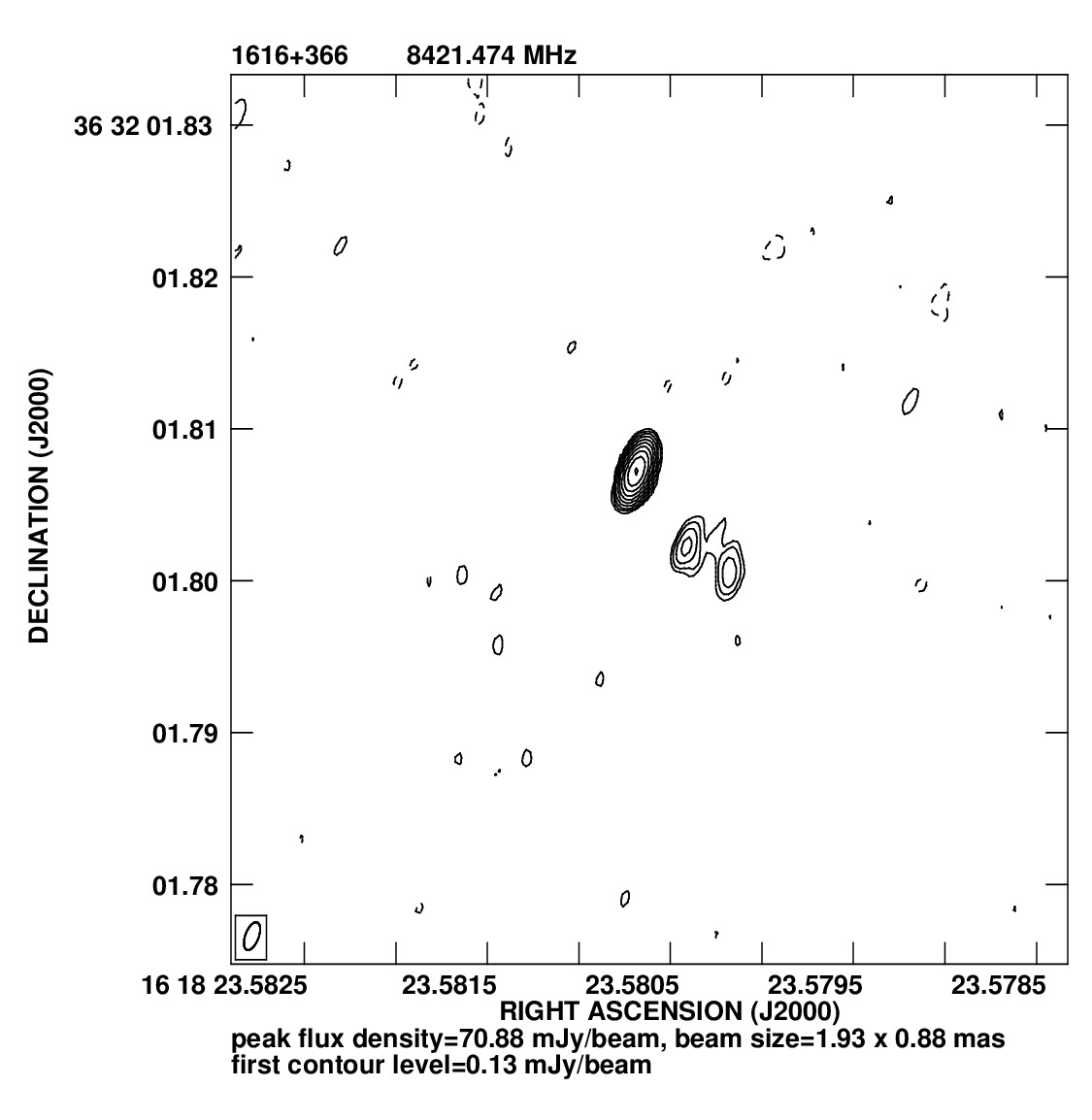}
\includegraphics[width=8cm, height=8cm]{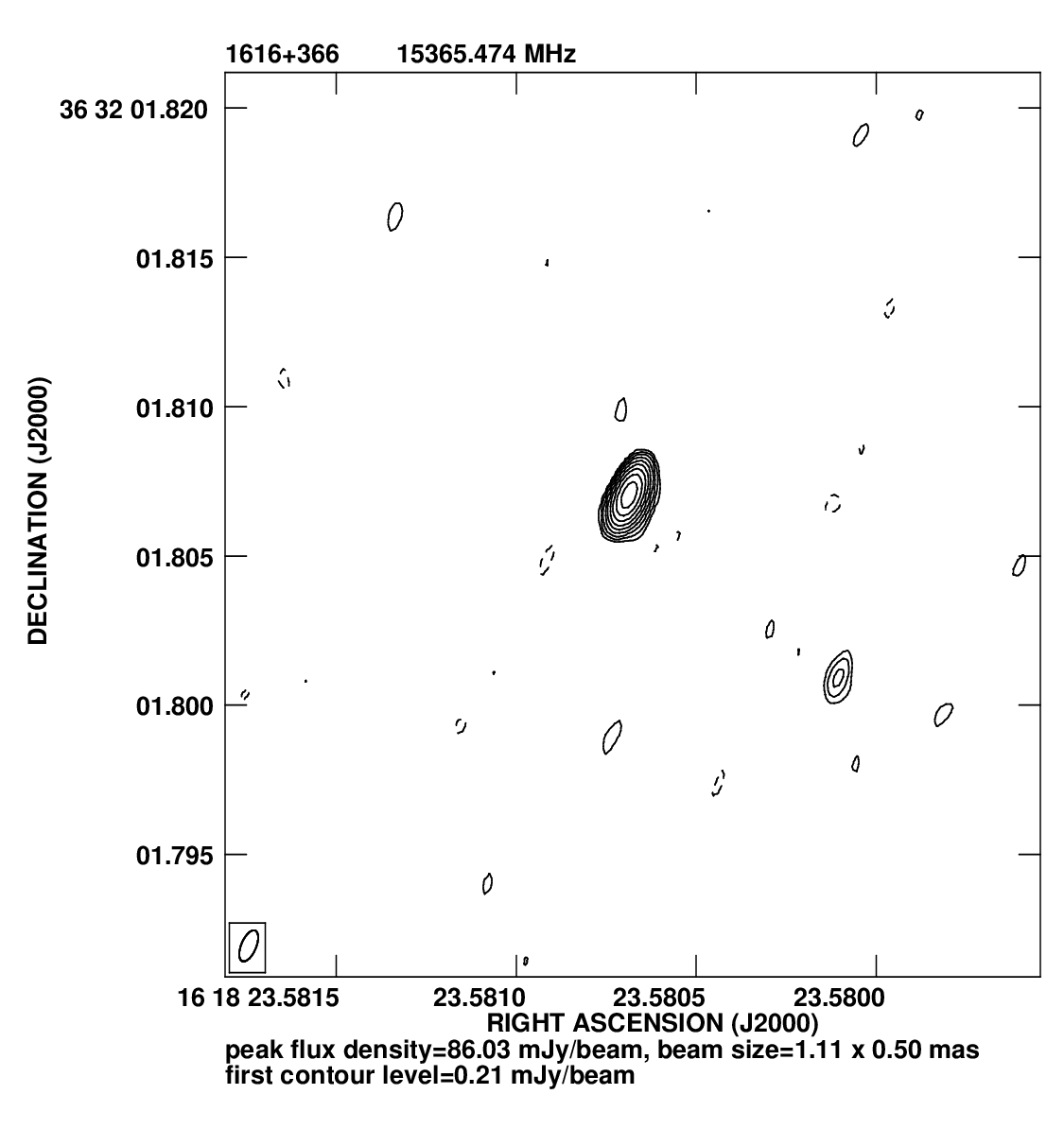}
\caption{The VLBA maps of 1616+366 at 1.65, 5, 8.4 and 15.4\,GHz.
Contours increase by a factor 2 and the first contour level corresponds
to $\approx 3\sigma$.}
\label{1616+366_maps}
\end{figure*}

\noindent {\bf \object{1616+366}}. A published VLA image of this radio galaxy
at 8.4\,GHz as a part of the Jodrell Bank--VLA Astrometric Survey 
\citep[JVAS,][]{pbww92} shows it to have a core-jet structure directed to
the south-west. The 1.65-GHz VLBA image (Fig.~\ref{1616+366_maps}) shows an
elongated structure that appears as a core-jet object at higher
frequencies. The radio core at the position RA=\,$16^{\rm h}18^{\rm m}23\fs581$,
Dec=\,$+36\degr 32\arcmin 01\farcs813$ has a flat spectrum between 1.65 and
15.4\,GHz. The interpolated total flux density at 1.65\,GHz is 488\,mJy, so our VLBA
image accounts for only $\sim$14\% of the total flux density. This percentage is
higher at 8.4\,GHz where our VLBA image accounts for $\sim$57\% of the total
flux density compared with the VLA image made by \citet{pbww92}.

1616+366 is present in SDSS/DR4 and has been identified as a galaxy with a
redshift of $z=0.734$ located at: RA=\,$16^{\rm h}18^{\rm m}23\fs582$,
Dec=\,$+36\degr 32\arcmin 01\farcs75$. 

\begin{table*}[t]
\caption[]{Flux densities of sources principal components at
observed frequencies.}
\begin{center}
\begin{tabular}{@{}c c c c c c c r r l@{}}
\hline
\hline
~~~Source & RA & Dec &
\multicolumn{1}{c}{${\rm S_{1.65\,GHz}}$}&
\multicolumn{1}{c}{${\rm S_{5\,GHz}}$}&
\multicolumn{1}{c}{${\rm S_{8.4GHz}}$}&   
\multicolumn{1}{c}{${\rm S_{15.4GHz}}$}&
\multicolumn{1}{c}{$\theta_{1}$}&
\multicolumn{1}{c}{$\theta_{2}$}&
\multicolumn{1}{c}{PA}~~~\\
~~~Name   & h~m~s & $\degr$~$\arcmin$~$\arcsec$ &
\multicolumn{1}{c}{mJy}&
\multicolumn{1}{c}{mJy}& 
\multicolumn{1}{c}{mJy}& 
\multicolumn{1}{c}{mJy}& 
\multicolumn{1}{c}{$\arcsec$}&
\multicolumn{1}{c}{$\arcsec$}&
\multicolumn{1}{c}{$\degr$}~~~\\
~~~(1)& (2) & (3) &
\multicolumn{1}{c}{(4)}&
\multicolumn{1}{c}{(5)}&
\multicolumn{1}{c}{(6)}&
\multicolumn{1}{c}{(7)}&
\multicolumn{1}{c}{(8)}&
\multicolumn{1}{c}{(9)}&
\multicolumn{1}{c}{(10)~~~}\\
\hline
~~~0809+404&08 12 53.123 &40 18 59.880 &287.6&~~193.9$\ast$ &~~~74.5$\ast$
                 &~~---&0.018&0.012&137~~~\\
        &08 12 53.124 &40 18 59.872 &288.9&  & 
                 &~~---&0.022&0.012&149~~~\\
\hline
~~~0949+287&09 52 06.102 &28 28 32.400 &199.8&~~29.6
                 &~~~8.5&~~---&0.015&0.007&~~83~~~\\
        &09 52 06.079 &28 28 32.417 &238.2&~~13.1
                 &~10.8&~~---&0.020&0.014&~~42~~~\\
        &--- &--- &~~238.5$\dagger$&~~~---
                 &~~~---&~~---&---&---&~~---~~~\\
\hline
~~~1159+395&12 01 49.964 &39 19 11.041 &271.2&~~86.5&~33.4
                 &~~---&0.008&0.006&172~~~\\ 
        &12 01 49.965 &39 19 11.003 &162.6&~~37.7 &~13.6
                 &~~---&0.007&0.006&~~52~~~\\
        &12 01 49.965 &39 19 11.028 &~~~---&~~~~5.1
                 &~~~---&~~---&---&---&~~---~~~\\
\hline
~~~1315+396&13 17 18.632&39 25 28.151&
                 ~~~1.5&~~~---&~~~---&~~---&0.013&0.006&105~~~\\
        &13 17 18.635&39 25 28.142&244.8&147.5&100.4
                 &51.9&0.005&0.001&106~~~\\
\hline
~~~1502+291&15 04 26.697 &28 54 30.554 &113.9&~~18.9 
                 &~~~---&~~---&0.005&0.003&~~27~~~\\
        &15 04 26.696 &28 54 30.545 &209.1&200.7
                 &~~~---&~~---&0.002&0.001&~~93~~~\\
        &15 04 26.698 &28 54 30.581 &~~32.1
                 &~~~---&~~~---&~~---&0.015&0.003&~~46~~~\\
        &15 04 26.705&28 54 30.494&~~12.3&~~~---&~~~---&~~---&0.014&
                 0.010 &126~~~\\
\hline
~~~1616+366&16 18 23.581 &36 32 01.813 &~~68.1 &~~67.1 &~77.1
                                    &95.6&0.014&0.003&~~46~~~\\
        &16 18 23.580 &36 32 01.802 &~~~---&~~20.7
                                    &~~~---&~~---&---&---&~~---~~~\\
\hline
\end{tabular}
\end{center}

\vspace{0.5cm} 
Description of the columns:
\begin{itemize}
\item[---]{Column~(1): Source name in the IAU format;}
\item[---]{Column~(2): Component right ascension (J2000) as measured at
1.65\,GHz;}
\item[---]{Column~(3): Component declination (J2000) as measured at
1.65\,GHz;} 
\item[---]{Column~(4): VLBA+Effelsberg flux density in mJy at 1.65\,GHz from
the present paper;}
\item[---]{Column~(5): VLBA flux density in mJy at 5\,GHz from the present
paper;} 
\item[---]{Column~(6): VLBA flux density in mJy at 8.4\,GHz from the present
paper;}
\item[---]{Column~(7): VLBA flux density in mJy at 15.4\,GHz from the
present paper;}
\item[---]{Column~(8): Deconvolved component major axis angular size at
1.65\,GHz obtained using JMFIT;}
\item[---]{Column~(9): Deconvolved component minor axis angular size at
1.65\,GHz obtained using JMFIT;}
\item[---]{Column~(10): Deconvolved major axis position angle at 1.65\,GHz
obtained using JMFIT;}
\end{itemize}
$\ast$ --- The flux density of the whole source at respective frequency
calculated by summing the map's clean components.\\
$\dagger$--- The flux density of diffuse matter of the eastern component
calculated by summing the respective clean components.

\label{table2}
\end{table*}

\section{Discussion}

\subsection{CSOs with arcsecond-scale relic extensions}

According to \citet{pol03} and references there\-in, CSOs are young radio
sources as their kinematic ages are of the order of $10^{3}-10^{4}$ years.
Recently, \citet{gug05} have investigated the ages of CSOs in a
systematic manner and have shown that there is a clear cutoff in the age
distribution at approximately 500~years, suggesting that CSOs may be young, not
only because they are in the initial stages in an evolutionary chain, but that
they are also short-lived i.e. their activity phase lasts for only a few
hundred years. The most straightforward cause of this is a lack of stable
fuelling. It follows that, because of a cutoff of the energy transport from the
core to the lobes, not only does the luminosity of the source drop, but also
diffuse radio lobes showing an absence of edge-brightening result as the
hotspots fade away quickly.

It could be that low frequency observations might reveal remnants of earlier
stages of activity as in the case of 0108+388, which is a compact double with
an arcsecond-scale relic extension $\sim$$20\arcsec$ east of the nucleus
\citep{b90}. \citet{ocp98} have shown that the bright component of 0108+388 is
a very young source with a kinematic age of 367\,years. To explain the diffuse
structure of 0108+388 they have adopted a scenario of recurrent activity in the
nucleus and have proposed an interpretation of the asymmetries of the extended
emission as being caused by light travel time effects. 

Contrary to this, \citet{b90} have suggested that 0108+388 could be a normally
aged radio galaxy in which most of the radio emitting plasma is unable to
escape from the nuclear region. Such a situation might arise if the host galaxy
has recently swallowed a gas rich companion, that has smothered the source.
The idea of a recent merger event is also a very plausible explanation of the
misalignment between the active (inner) and inactive (outer) parts of the
source.

0108+388 is the first known example of a CSO with an arcsecond-scale structure.
Moreover, it has also been classified as a GPS source because of a spectral
turnover, which, according to \citet{marr01}, results from free-free
absorption by nonuniform gas, possibly in the form of a disk in the central
tens of parsecs. Instabilities in such a disk could result in a periodic infall
of gas, that would produce apparent renewed activity. 
\citet{car98} have also found significant H\,I absorption along the line of
sight to the core of 0108+388, suggesting the existence of a large amount of
thermal gas.

Another example of a CSO possessing a (relatively) large-scale structure is
0402+379 \citep{man04}. It has an arcsecond-scale, core-dominated structure
and has been classified as an MSO. Unlike 0108+388, the outer lobes of
0402+379 are symmetric.

\subsection{The case of 0809+404}\label{s-0809} 

As the VLA images of
0809+404 (F2001) show it to have a very asymmetric double structure with
1\farcs2 separation and a flux density ratio $\sim$100:1 at 4.9\,GHz, we
suggest that, as far as the arcsecond structure is concerned, 0809+404
resembles 0108+388. The weaker western component of the 0809+404 VLA structure
is a relic of previous activity and its spectral index, calculated from the
4.9 and 8.5\,GHz VLA observations, is very steep ($\alpha=-1.5$). This component
is not present in the 15-GHz VLA image. There is also no
indication of a radio core located elsewhere in this image.
However, on a milliarcsecond scale, the differences
between the images of 0108+388 \citep[see e.g.][]{tay96}, and ours of 0809+404
become apparent: 0108+388 is a triple with mini-FR\,II-like lobes and a core,
whereas no clear FR\,II-like structure is present in 0809+404, although there
is a hint of the existence of two lobes in the 1.65-GHz image.
Neither a core nor hotspots are observable at any frequency and the whole
milliarcsecond-scale structure fades away towards higher frequencies, indicating
that activity in the nucleus has switched off.

It can be seen that there is an appreciable
($\sim$$40\degr$) misalignment between the axis of the inner structure with
respect to
the outer, relic structure. Perhaps, as in the case of 0108+388, a recent
merger event in 0809+404 is also a likely scenario.

The eastern bright component shows little Faraday rotation but is strongly
depolarized \citep[F2001,][]{fanti04}. 
The polarization asymmetry is very common among small-scale CSS
sources and can be caused by differences in the gas density in the
surrounding medium \citep{tsm03}. We have begun a programme using the WSRT to
investigate the gaseous medium of 23 sources from the parent sample (Paper~I)
with known redshifts. So far three of them have been observed during WSRT
service time at UHF-high frequencies (700--1200\,MHz) and 0809+404 was 
one of them. Approximately 3 hours of
observing time were spent on each object, but none of them has shown any H\,I
absorption. For 0809+404, a 2$\sigma$ upper limit of $\tau$$\sim$0.005 for
the optical depth has been set based upon a noise level of
$\sim$3.7\,mJy/beam. The results of these observations will be described in
detail in a future paper.

\citet{stan03} and \citet{stan05} have discussed in detail the properties of
a few GPS objects with extended arcsecond-scale emission and have shown
that the extended emission around CSO/GPS galaxies is well explained as the
co-presence of past and new activity. GPS quasars without a CSO
morphology, but having arcsecond-scale emission, are likely to be intrinsically
large, old and active radio sources seen along the radio axis, whereas CSO
morphologies in quasars, or at least mini-lobe dominated
structures, may be signatures of a small and young radio source. 

0809+404 is classified spectroscopically as a quasar in SDSS/DR4. More
precisely, \citet{zakam03} have found this object to be one of 291 type-II
quasar candidates. The spectra of these objects are
dominated by narrow emission lines, so their broad-line emission region as
well as the UV-continua are completely obscured at optical wavelengths by a
dusty torus. It can be assumed that the axes of such objects are perpendicular
to the line of sight. This assumption and the fact that the
integrated spectrum of 0809+404 is not of a GPS type (M.~Murgia, priv.
comm.) is in agreement with the conjecture that this object is not
beamed to the observer. Besides, based on his findings, it is plausible that
the inner structure of 0809+404 is younger than the arcsecond-scale emission.

The radio spectrum of 0809+404 is well described by a continuous injection (CI)
model (which refers to the source as a whole) with $\alpha_{inj}=-0.4$ and a
break frequency $\nu_{br}$= 1.56\,GHz (M.~Murgia, priv. comm.),
suggesting that the radio source is continuously replenished by a constant flow
of fresh relativistic particles. The initial spectral index $\alpha_{inj}$,
which is the spectral index of the synchrotron radiation in the part of the
spectrum not affected by the evolution, implies a power law for the energy
distribution of the injected electrons of $\delta\approx$1.8. During the CI
phase, the electrons lose energy by synchrotron emission and inverse Compton
scattering of the cosmic microwave background photons.

The continuous injection model does not contradict the
possibility of intermittent activity, since, during the lifetime of the extended
radio emission, the nucleus appears ``on average'' active.
Assuming minimum energy conditions we have estimated a source age
$t_{sync}\sim 10^{5}$ years using the formulae from \citet{miley}
(eq.\,[7]), where $\theta_{x}$ and $\theta_{y}$ correspond to the beam widths
of the 4.9-GHz VLA map (produced from the data extracted from the VLA archive).
The minimum
energy magnetic field was calculated by using eq.\,[2] from \citet{miley} and
the standard assumptions about the lower and upper cutoff frequencies (10\,MHz
and 100\,GHz respectively), the uniform filling factor ($\eta=1$), the
ratio between protons and electrons ($k=1$), the angle between the magnetic
field and the line of sight (90\degr) and an equivalent field of the cosmic
background radiation ($B_{CMB}=3.25(1+{\it z})^2~[{\rm \mu G}]$). The
calculations were made for 4.9\,GHz and the corresponding flux density was
taken from F2001.

A number of theories trying to explain the episodic activity of radio sources
exist in the literature. According to one, activity could be initiated as a
result of a merger event. Torques and shocks during the merger can remove
angular momentum from the gas in the merging galaxies and this provides
injection of substantial amounts of gas/dust into the central nuclear regions
\citep{mh96}. It is, therefore, likely that in the initial phase of an AGN,
this gas still surrounds (and possibly obscures) the central regions. Such
activity can perhaps cease when there is no more matter to be accreted.

\citet{tin03} have also suggested that at least some GPS sources are limited
in their development by the effects of merger activity and the resulting likely
sporadic fuelling of the central black holes and accretion disks. \citet{bar96},
who tracked the evolution of both gas and stars in the merger of two disk
galaxies, show that in the final state of such a merger, 60\% of the gas is
driven into the inner part of the galaxy to within 100\,pc of the nucleus. In
such an environment, a black hole may undergo many fuelling events and each
event may completely disrupt and/or restart the jet. After each renewal, the
jet may need to force its way through the dense nuclear environment anew.

Alternatively, to interpret the nature of the inner structure of 0809+404 a
theory of the Super\-mass\-ive Black Hole (SMBH) accretion disk instabilities
\citep[][and references therein]{hat01,jan04} could be adopted. It takes into
account a non-stationary accretion due to hydrogen ionization that can
develop in the disk. These instabilities can cause a sudden accretion of the
material surrounding the source and accumulated in the outer torus. It also
predicts that galaxies spend the greater part of their lifetime, say $\sim$70\%,
in a ``quiescent'' state and $\sim$30\% in an active state with the length of
the active phase of an AGN as well as the timescale of the re-occurrence of
activity being determined by the mass of the SMBH. Specifically, if the SMBH
mass is assumed to be of the order of $10^7$ M$_\odot$ -- and such an
assumption is plausible for RLAGNs \citep{wo02,osh02} -- the length of the
activity period may be as low as $\sim$$10^3$ years. This means that the
transition to the fader phase can also happen at a very early stage of
evolution i.e. at the CSO stage. Note also that such an interpretation is in
agreement with an early hypothesis on the episodic nature of CSOs given by
\citet{r94} and recent findings of \citet{gug05}.
However, there is some inconsistency if such an interpretation is applied to
0809+404, namely that a misalignment between the inner and outer parts of the
source is observed. This indicates that the previous period of activity may
have been linked to a recent merger. It is to be noted that misalignments
between the inner and outer parts of the source can be observed regardless of
the source scale. 0809+404 and 0108+388 are good examples of very compact
sources accompanied by arcsecond-scale relics. In 1245+676, the inner
CSO part shows a modest misalignment with respect to the outer megaparsec-scale
structure \citep{mar16} whereas in a few core-dominated sources shown by
\citet{mtm06} large misalignments of the kiloparsec-scale structures with
respect to the outer large-scale ones are quite common.

Another interpretation of the
nature of this source is also conceivable, namely that the
brighter eastern component of the arcsecond-scale structure is a radio lobe.
However, even in this scenario, the lack of a visible hotspot in the eastern,
dominating component clearly suggests that the source as a whole is a fader.

\subsection{The other five sources}

The morphology of the edge-brightened lobes of 0949+287 as seen at 1.65\,GHz
suggests this source is an FR\,II like object, with a steep integrated spectrum
\citep{bol04}. To calculate its linear size and power we
adopted a redshift of $z=1.085$, which is the median value of all the available
redshifts of the 60 sources from our primary sample. The estimated linear size
and the calculated power at 1.4\,GHz (Table~\ref{table1})  
indicate that 0949+287 could be a bright CSS object.

1159+395 is a highly redshifted galaxy with a double structure.
It is the most powerful object in our sample (Table~\ref{table1}).
It appears to be unresolved in the VLA images at 4.9 and 8.5\,GHz (F2001) so it
can be assumed that it has no extended emission. Its
double structure with compact features identified as hotspots as well as a
non-GPS spectrum \citep{mur99} suggest that it is a typical ``active'' CSO.
According to \citep{mur99}, 1159+395 is a very young source with a synchrotron
age of the order of $\sim10^3$\,years.

The remaining three sources from the subsample presented here (1315+396,
1502+291, 1616+366) have very compact core-jet structures clearly indicating
that these sources are not CSOs so that, even if potentially they could be
intermittent, they are in an active phase at the current epoch.

\section{Summary}

Multifrequency VLBA observations of six highly compact yet steep spectrum
objects from our parent sample of weak CSS sources (Paper~I) have been made.
The observations presented here (as well as those of Paper~II) suggest
that some CSS and CSO sources can be short-lived objects. Our results presented
so far indicate that fading sources are rare among compact objects with a radio
power of order of $10^{26}$ -- $10^{28}{\rm W~Hz^{-1}}$ at 1.4\, GHz. However,
given that small-scale objects are likely to be overpressured \citep{siem05},
the expansion losses would dim a compact source {\em quickly} once the central
engine switches off. Therefore, such sources are inevitably weaker than their
respective ``still active'' counterparts and, consequently, have largely
escaped detection in the GHz-frequency range because of their low
radio power. Furthermore, the fader stage of a compact source is rather
ephemeral and hence not easily observable. Nevertheless, this selection effect
does not preclude the phenomenon of a premature switch-off of a RLAGN from
being a direct cause of the small-scale source excess confirmed either
statistically \citep{fanti90,odea97} or observed directly \citep{gug05}.

Apart from a natural weakening of the sources, the early cutoff of the energy
transport
via jets shortly makes the lobes take the form typical for faders, albeit
without their spectra showing signs of ageing for frequencies below 5\,GHz.
Therefore, further investigations of LPC sources, i.e. those with radio powers
below $\sim$$10^{26}{\rm W~Hz^{-1}}$ at 1.4\,GHz and sub-arcsecond angular sizes
\citep{gir05}, but not necessarily ultra-steep spectra, might lead to the
discovery of more members of compact fader class.

In this context, 0809+404 appears to be an object of a particular importance. It
is a young source with a synchrotron age of the order of $10^{5}$ years, which
is typical for CSS objects. However, its morphological structure suggests a
more complex past. The weak western component in the VLA image is possibly a
relic of previous activity. The bright eastern component appears amorphous with
no hotspots in the milliarcsecond scale. Two explanations for the unusual
features observed in the VLA (F2001) and our VLBA images are plausible:

\begin{enumerate}
\item Restarted activity as an outcome of a merger, shortly followed by a decay
of the structures resulting from the new period of activity.
\item Very asymmetric arcsecond-scale double source that is fading away.
\end{enumerate}

Based upon the morphological structures of 0809+404 seen in the VLA and VLBA
images and H\,I absorption observations, we are leaning to the first
interpretation and consider the arcsecond-scale weak western component,
analogous to that in 0108+388, as a signature of a past epoch of activity
whereas the dominating eastern component is likely to be a very compact fader.
The latter interpretation is consistent with the observational evidence
brought by \citet{gug05} suggesting that such short periods of
activity in AGNs are possible. 

Particularly useful would be a high resolution optical image with the
sensitivity enabling the search for the evidence of a merger or its aftermath.
This could lead to an explanation of the nature of the weak
western component. The currently available optical image of 0809+404 from
SDSS/DR4 is not helpful with regard to this.

\begin{acknowledgements}

\item The VLBA is operated by the National Radio Astronomy Observatory
(NRAO), a facility of the National Science Foundation (NSF) operated under
cooperative agreement by Associated Universities, Inc. (AUI).

\item Effelsberg telescope is operated by the
Max-Planck-Institut f\"ur Radio\-astronomie (MPIfR) and it is a part of the
European VLBI Network (EVN). 

\item This research has made use of the NASA/IPAC Extragalactic Database
(NED) which is operated by the Jet Propulsion Laboratory, California
Institute of Technology, under contract with the National Aeronautics and
Space Administration.

\item Use has been made of the third release of the Sloan Digital Sky
Survey (SDSS) Archive. Funding for the creation and distribution of the
SDSS Archive has been provided by the Alfred P. Sloan Foundation, the
Participating Institutions, the National Aeronautics and Space
Administration, the National Science Foundation, the U.S. Department of
Energy, the Japanese Monbukagakusho, and the Max Planck Society. The SDSS
Web site is http://www.sdss.org/. The SDSS is managed by the Astrophysical
Research Consortium (ARC) for the Participating Institutions. The
Participating Institu\-tions are The University of Chicago, Fermilab, the
Insti\-tute for Advanced Study, the Japan Participation Group, The Johns
Hopkins University, Los Alamos National Labora\-tory, the
Max-Planck-Institute for Astronomy (MPIA), the Max-Planck-Institute for
Astrophysics (MPA), New Mexico State University, University of Pittsburgh,
Princeton University, the United States Naval Observatory, and the
University of Washington.

\item We thank Raffaella Morganti for her help with the WSRT service time
observations.

\item We thank Karl-Heinz Mack for reading of the early version of this paper
and a number of suggestions.

\item This work was supported by Polish Ministry of Education and Science under
grant 1 P03D 008 30.

\end{acknowledgements}

\end{document}